\documentclass{aa}
\usepackage{natbib}
\usepackage{txfonts}
\usepackage{hyperref}
\usepackage{graphicx}
\graphicspath{{Fig/}}

\newcommand{\grad}{\vec\nabla}

\newcommand{\lp}{ \;\left(}
\newcommand{\rp}{ \right)}
\newcommand{\nab}{ \vec{\nabla} }
\newcommand{\n}{\nabla }
\newcommand{\lapl}{ \Delta }

\newcommand{\beq}{\begin{equation}}
\newcommand{\eeq}{\end{equation}}

\newcommand{\lc}{ \left[}
\newcommand{\rc}{ \right]}
\newcommand{\greq}{\begin{equation} \begin{array}{l}}
\newcommand{\egreq}{\end{array} \end{equation}}

\newcommand{\eeqn}[1]{\label{#1}\end{equation}}

\newcommand{\eq}[1]{(\ref{#1})}

\newcommand{\beqa}{\begin{eqnarray}}
\newcommand{\eeqa}{\end{eqnarray}}
\newcommand{\beqan}{\begin{eqnarray}}
\newcommand{\eeqan}[1]{\label{#1}\end{eqnarray}}

\begin{document}
 
\title{Asymptotic theory of gravity modes in rotating stars. I.~Ray dynamics}

\author{V. Prat \inst{1,2,3} \and F. Ligni\`eres \inst{2,3} \and J. Ballot \inst{2,3}}
\institute{
Max-Planck Institut f\"ur Astrophysik, Karl-Schwarzschild-Str. 1, 85748, Garching bei M\"unchen, Germany\\
\email{vprat@mpa-garching.mpg.de} \\
\and
Universit\'e de Toulouse; UPS-OMP; IRAP; Toulouse, France \\
\and
CNRS; IRAP; 14 avenue \'Edouard Belin; F-31400 Toulouse, France}

\offprints{F. Ligni\`eres}

\date{}

\abstract
{The seismology of early-type stars is limited by our incomplete understanding of gravito-inertial modes.}
{We develop a short-wavelength asymptotic analysis for gravito-inertial modes in rotating stars.}
{The Wentzel-Kramers-Brillouin approximation was applied to the equations governing adiabatic small perturbations
about a model of a uniformly rotating barotropic star.}
{A general eikonal equation, including the effect of the centrifugal deformation, is derived.
The
dynamics of axisymmetric gravito-inertial rays is solved numerically for polytropic stellar models of increasing rotation
and analysed by describing the structure of the phase space.
Three different types of phase-space structures are distinguished.
The first type results from the continuous evolution of structures of the non-rotating integrable phase space.
It is predominant in the low-frequency region of the phase space.
The second type of structures are island chains associated with stable periodic rays. 
The third type of structures are large chaotic regions that can be related to
the envelope minimum
of the Brunt-V\"{a}is\"{a}l\"{a} frequency.}
{Gravito-inertial modes are expected to follow this classification, in which the frequency spectrum is a superposition of
sub-spectra associated with these different types of
phase-space structures.
The detailed confrontation between the predictions of this ray-based asymptotic theory and numerically computed modes
will be presented
in a companion paper.}


\maketitle

\section{Introduction}

As compared to the wealth of new information obtained from the seismology of solar-type and red-giant stars  \citep{M15},
reliable seismic diagnostics on non-evolved, intermediate-mass and massive pulsators ($\gamma$ Doradus, $\delta$ Scuti, $\beta$ Cephei, SPB
or Be stars)
are scarce and, in any case, restricted to atypical slowly rotating stars \citep{KS14,SK15,TM15}.
This stems from our limited understanding of the effects of rotation on oscillation modes \citep{L13,T14,R15}, at least for the high rotation rates of
typical non-evolved, intermediate-mass and massive stars \citep{RZ07,L13}.
The two classical approximations to treat rotational effects have been the perturbative expansion in $\Omega/\omega$, where
$\Omega$ is the rotation rate and $\omega$ is the pulsation frequency \citep{S81}, and the so-called traditional approximation
of the Coriolis acceleration \citep{E60}. The perturbative expansion happens to be valid for small rotation rates but not for those 
of most early-type stars \citep{RL06, BL10, BL13}. 
The traditional approximation greatly simplifies the understanding of Coriolis effects on low-frequency g modes \citep[e.g.][]{T03, BD13} but
the range of validity and  accuracy of this approximation are not known. The centrifugal deformation in particular is not taken into account 
in the traditional approximation.

More recently, the development of dedicated numerical codes has made it possible to compute  
modes and frequencies accurately without approximating the Coriolis and centrifugal effects on the modes \citep{RL06,Reese09,OD12}. This is a crucial step
towards a confrontation
with observational data. However, in addition to accurate theoretical frequencies, mode identification 
of 
observed frequencies requires \emph{a priori} knowledge 
of the mode properties.
In the case of slowly rotating stars, our knowledge of the expected frequency patterns is important for the interpretation of observed frequency spectra.
It comes from 
the
short-wavelength asymptotic theory of oscillation modes, which
in the case of
spherically symmetric non-rotating stars provides analytical expressions of the asymptotic frequency patterns.
When applied to rapidly rotating stars, the same short-wavelength asymptotic analysis is not as straightforward. 
As in geometrical optics, it first leads to a ray model 
that describes wave propagation. Then, the conditions to produce a mode from 
the positive interference between rays that need to be established.
Methods and concepts to infer the properties of the modes from those of the rays have been mostly developed in the field of quantum physics.
The basic idea is to study rays as trajectories of a Hamiltonian
system and then to relate the modes to the phase-space structures of the ray dynamics.
For acoustic modes in rotating stars, such a ray-based asymptotic theory provided a physical classification of the modes
as well as quantative predictions that
have been successfully confronted with numerically computed high-frequency p modes \citep{LG1, LG2, PG11, PL12}.

In this paper, we intend to construct a similar ray-based asymptotic theory for gravito-inertial modes. 
Equations governing gravito-inertial rays in rotating stars are derived and then solved to obtain a global view of the properties of axisymmetric gravito-inertial rays and their evolution with stellar rotation.
The detailed confrontation with numerically computed modes
will be presented
in a companion paper, although a first successful comparison in the particular case of so-called rosette modes has been already presented in a proceedings
paper \citep{BL12}.

Ray models are usually obtained from the Wentzel-Kramers-Brillouin (WKB) approximation of the equations governing small-amplitude perturbations,
and this approximation is valid when
the wavelength of the perturbation is much shorter than the length scale of the background spatial variations. 
Gravito-inertial ray models
have been first developed for the ocean and the Earth's atmosphere. 
One of the main motivation in this context is to model angular momentum transport and deposition by waves \citep{FA03,ME95,BR04}.
In stellar physics, such WKB ray models have been limited to pure gravity waves \citep{G93,AS15} and have not been used to study
angular momentum transport or gravito-inertial modes.
Instead, this is another ray theory, the method of characteristics,
which has been applied by \citet{DR00} to the problem of stellar gravito-inertial waves and their relation with gravito-inertial modes. As
detailed in \citet{HM06}, the WKB approximation and the method of characteristics 
differ in general, although they can be identical 
in some specific cases. 
For what concerns stellar gravito-inertial waves, they differ in that
the method of characteristics cannot take into account the compressibility effects that produce the back-refraction of the waves approaching the stellar surface.
We thus expect WKB ray models to be more realistic than the method of characteristics.

The paper is organised as follows. In Sect.~\ref{sec:eikonal}, the WKB approximation is applied to the wave equation governing small perturbations
about a model of a uniformly rotating star. A general eikonal equation for gravito-inertial waves is derived and the domains of propagation 
of these waves inside the star are discussed.
In Sect.~\ref{sec:raymodel}, the ray model for axisymmetric rays is derived from the eikonal equation, its Hamiltonian nature is emphasised and the standard tool to investigate the phase-space 
structure of
the Hamiltonian ray dynamics is
presented. 
A detailed numerical study of the ray dynamics for uniformly rotating polytropic models of stars is then presented in Sect.~\ref{sec:dynamics}.
The interpretation of the dynamics in terms of gravito-inertial modes is discussed in Sect.~\ref{sec:modes} and  conclusions are provided in Sect.~\ref{sec:discussion}.

\section{Eikonal equation for gravito-inertial waves}
\label{sec:eikonal}

In this section, we derive an eikonal equation for short-wavelength, adiabatic gravito-inertial waves in a 
uniformly rotating polytropic model of star.
The stellar model is specified in Sect.~\ref{sec:poly}, the equations for small perturbations about this model are put in a convenient form in Sect.~\ref{sec:perturb},
the eikonal equation obtained by applying the WKB approximation to the perturbation equations is presented in Sect.~\ref{sec:wkb}, and 
the domains of propagation of gravito-inertial waves are
discussed in Sect.~\ref{sec:domain}. Only the main equations are presented in this section because their derivation are detailed in Appendix~\ref{sec:deriv}.

\subsection{Polytropic model of rotating star}
\label{sec:poly}

The stellar model is
a self-gravitating uniformly rotating monatomic gas that  verifies a polytropic relation, which is
assumed to give a reasonably
good approximation
of the relation between pressure and density in real stars \citep{HK94}.
The governing equations can be written as
\begin{eqnarray} \label{hyd}
P_0 &=& K \rho_0^{1+1/\mu}, \\
0 &=& -\nab P_0 - \rho_0 \nab \left( \psi_0 -\Omega^2 w^2/2 \right),  \\
\lapl \psi_0 &=& 4\pi G\rho_0,
\end{eqnarray}
where $P_0$ is the pressure, $\rho_0$ the density, $K$ the polytropic constant,
$\mu$ the polytropic index, $\psi_0$ the gravitational potential, $\Omega$ the rotation rate,
$w$ the distance to the
rotation axis, and $G$ the gravitational constant.

The uniform rotation ensures that the
flow is barotropic.
A pseudo-enthalpy $h_0=\int dP_0/\rho_0 = (1+\mu)P_0/\rho_0$ can then be introduced and quantities
describing the equilibrium model, such as the sound speed $c_{\rm s} = \sqrt{{\Gamma_1} P_0 / \rho_{0}}$, where $\Gamma_1=5/3$ is the first adiabatic exponent of the gas;
the effective gravity $\vec{g}_0 = - \nab \left( \psi_0 -\Omega^2 w^2/2 \right)$; and the Brunt-V\"{a}is\"{a}l\"{a} frequency, given by the relation $N_0^2= \vec{g}_0 \cdot \lp \frac{\nab \rho_{0}}{\rho_{0}} - \frac{1}{\Gamma_1}\frac{\nab P_0}{P_0} \rp$ 
 are expressed in terms of $h_0$:
\beqa
\label{eq:enth}
\vec{g}_0 = \nab h_0, \qquad \;  c_{\rm s}^2 = \frac{\Gamma_1}{\mu+1} h_0, \qquad \; N_0^2 = \lp \frac{\mu \Gamma_1}{\mu+1} -1 \rp \frac{g_0^2}{c_{\rm s}^2}.
\eeqa
In polytropic models of stars, the density $\rho_0$ vanishes at the surface.
Moreover, the polytropic relation Eq.~\eqref{hyd} implies that the sound speed $c_{\rm s}$ also vanishes, and the last of Eqs.~\eqref{eq:enth} shows that the Brunt-V\"ais\"al\"a frequency $N_0$ goes to infinity.
These properties are not necessarily valid in more realistic stellar models.

Specifying the mass and rotation rate of the star is not sufficient to determine
the polytropic model in physical units.
This requires fixing an additional parameter, for example the stellar radius 
(suitable parameter choices are discussed  in \citet{HK94}, for the non-rotating case, and 
in \citet{CT99}, for the rotating case).
In the following, however, 
we only present dimensionless quantities that do
not depend on the choice of this additional parameter.
The rotation rate $\Omega$ is compared to $\Omega_{\rm K} = \left(GM/R_{\rm e}^3\right)^{1/2}$,
the limiting rotation rate for which the centrifugal acceleration equals the gravity at the equator, where $M$ is the stellar mass and $R_{\rm e}$ the equatorial
radius. Other frequency units important for our problem are 
(i) the
inverse of the dynamical timescale associated with hydrostatic equilibrium $\omega_0 = \left(GM/R_{\rm p}^3\right)^{1/2}$, which provides a lower bound for acoustic wave frequencies, where $R_{\rm p}$ is the polar radius;
(ii) the Brunt-V\"{a}is\"{a}l\"{a} frequency $N_0$, an upper bound for gravity wave frequencies; and (iii) twice
the rotation rate $f=2\Omega$, the upper bound for pure inertial wave frequencies.

\begin{figure}
\resizebox{\hsize}{!}{\includegraphics{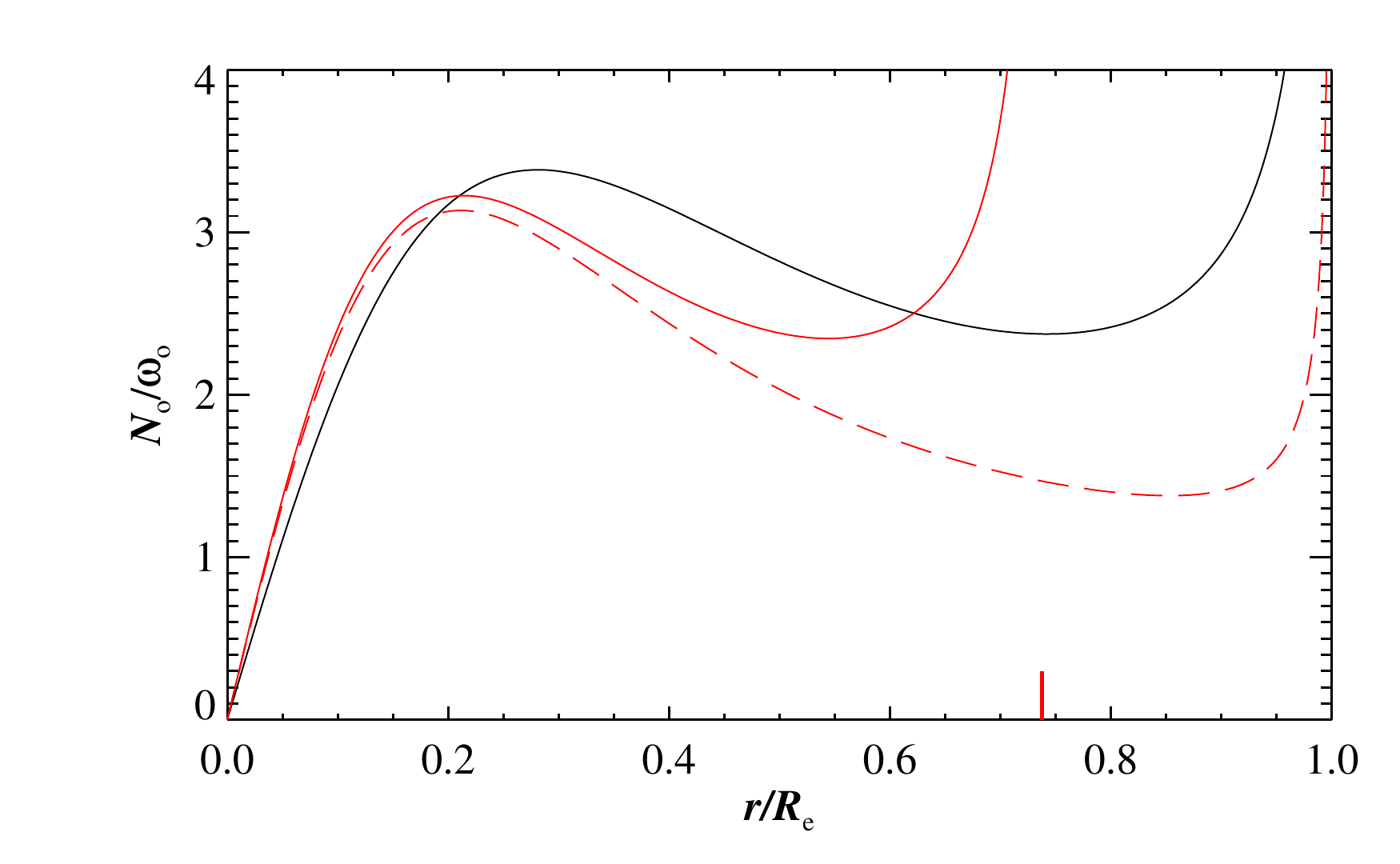}}
\resizebox{\hsize}{!}{\includegraphics{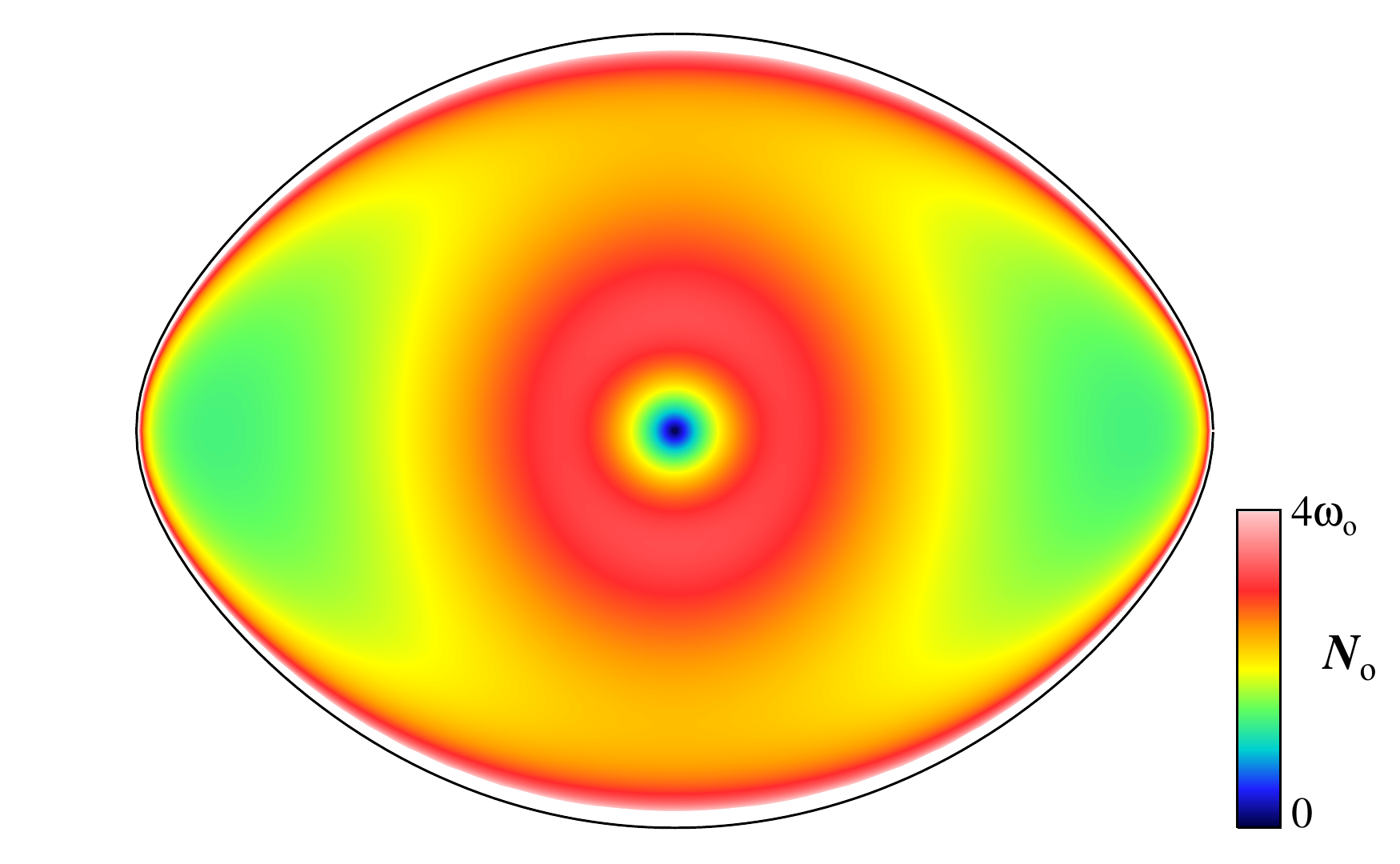}}
\caption{
(Top)
Solid black line shows
the profile of the Brunt-V\"{a}is\"{a}l\"{a} frequency $N_0$
normalised by $\omega_0$
for the non-rotating stellar model. The red lines
show
$N_0$ for the $\Omega = 0.84 \Omega_{\rm K}$ model 
along the polar (solid line) and equatorial (dashed line) radii. A thick red tick on the
x-axis indicates the polar radius $R_{\rm p}$.
(Top)
Map in a meridional plane of
$N_0$ normalised by $\omega_0$ for the $\Omega = 0.84 \Omega_{\rm K}$  model. 
The black line is the star surface.}
\label{Brunt}
\end{figure}

The spatial distribution of the Brunt-V\"{a}is\"{a}l\"{a} frequency is shown in Fig.~\ref{Brunt}. In the non-rotating spherically symmetric case (see the black curve on the upper figure), 
$N_0$ 
vanishes at the centre, has a local maximum, $N_{0,{\rm max}}$, around $r= 0.28 R$ and then shows a local minimum, $N_{0,{\rm min}}$, in the stellar envelope near $r = 0.74 R$.
A simple analytical expression of $N_0$, which is  valid for envelope models, (see Appendix~\ref{sec:hyperbolic}) shows that a local minimum is present at $r=3/4 R$ for all
polytropic indices, suggesting that such a minimum of $N_0$ is a generic feature of stars with radiative envelopes.
The Brunt-V\"{a}is\"{a}l\"{a} frequency distribution becomes anisotropic in centrifugally deformed stars. In particular, 
the amplitudes and positions of the local minima are clearly different 
along the polar and the equatorial axes.
This is apparent in Fig.~\ref{Brunt}, which shows iso-contours of $N_0$ (bottom) as well as the polar and equatorial radial profiles of $N_0$ (top)
for a $\Omega/\Omega_{\rm K} = 0.84$ model. In the following,
the local minima along the polar and equatorial axes are denoted $N^{\rm p}_{0,{\rm min}}$ and $N^{\rm e}_{0,{\rm min}}$, respectively.
By comparison, the variation of the local maximum of the Brunt-V\"{a}is\"{a}l\"{a} frequency between its values along the pole ($N^{\rm p}_{0,{\rm max}}$) and along
the equatoral radius ($N^{\rm e}_{0,{\rm max}}$)
remains small. This is 
because the deviation from sphericity 
is weaker in the central layers.

\subsection{Perturbation equations and boundary conditions}
\label{sec:perturb}

Time-harmonic, small-amplitude perturbations of the stellar model are studied under two
approximations. The first approximation is to neglect 
viscous and thermal dissipations. Both processes have a very small effect on the value of oscillation frequencies, at least in the frequency range of
observed 
modes. Non-adiabatic processes play an essential role in the oscillation amplitude, in particular, through excitation mechanisms such as the $\kappa$ mechanism,
but determining mode amplitudes is outside the scope
of the present study.
The second approximation, known as the Cowling approximation, is to neglect perturbations of the gravitational potential induced by density perturbations. 
It is justified for oscillations of small wavelengths and, therefore, fully compatible with the small-wavelength regime considered here.
Under these two approximations, the linear equations governing the evolution
of small-amplitude perturbations can be written as

\begin{align}
{\partial}_t \rho + \nab \cdot (\rho_{0} \vec{v})               &= 0, \label{div}        \\
\rho_{0} {\partial}_t \vec{v} + \rho_{0} \vec f\wedge\vec{v}    &= - \nab P + \rho \vec{g}_0,  \label{vel} \\
{\partial}_t P + \vec{v} \cdot \nab P_0                         &= c_{\rm s}^{2} \lp {\partial}_t \rho + \vec{v} \cdot \nab \rho_0 \rp, \label{adia}
\end{align}
where $\vec{v}$, $\rho$, and $P$, are respectively the Eulerian perturbations of velocity,
density, and pressure, and $\vec f= 2 \vec \Omega$ is the rotation vector.

Before applying the short-wavelength approximation, the perturbation equations for $\vec{v}$, $\rho$, and $P$ are reduced to a single equation for $\hat{P}$,
the complex amplitude of the time-harmonic pressure perturbation $P = \Re\lbrace \hat{P}(\vec x) \exp (- i \omega t)\rbrace$. According to the 
calculations detailed in Appendix~\ref{sec:pressure}, we obtain the equation
\beqa
\label{good1}
\Delta\hat P = \frac{f^2}{\omega^2}\n_{z}^2\hat P + \frac{N_0^2}{\omega^2} \Delta_{\perp}\hat P + (\vec{\cal V}' + i {\vec{\cal V}'}_{\!\!\!\!m}) \cdot\grad \hat P + M' \hat P,
\eeqa
where the operators $\n_{z}^2$ and $\Delta_{\perp}$ are related to the unit vectors parallel to the rotation axis, $\vec {e}_z =\vec{f}/f$, and the effective gravity, $\vec e_{\parallel} =-\vec g_0 /g_0$, by
\beqa
\n_{z}^2  \equiv  \n_{z} (\n_{z} ) &   \mbox{with} \;\;\;\; \n_{z} \equiv \vec {e}_z  \cdot \nab, \label{gr0}\\
\Delta_{\perp}  \equiv  \grad\cdot(\grad_{\perp}) & \mbox{with} \;\;\;\; \grad_{\perp} \equiv \grad - \vec e_{\parallel}  \n_{\parallel}  \;\;\;\; \mbox{and} \;\;\;\; \n_{\parallel} \equiv \vec e_{\parallel}  \cdot \nab. \label{gr00}
\eeqa
The expressions of $\vec{\cal V}', {\vec{\cal V}'}_{\!\!\!\!m}$, and $M'$ are given by Eqs.~(\ref{v0}, \ref{vm}, and \ref{mm}), respectively. 
As a linear
combination of $\vec{f}$ and $\vec g_0 $, the vector $\vec{\cal V}'$ 
is 
contained in a
meridional plane, 
whereas ${\vec{\cal V}'}_{\!\!\!\!m}$ is colinear to $\vec{f}\wedge\vec g_0 $ and thus perpendicular to this plane.
First order 
derivatives in the meridional plane can be suppressed
from this equation by introducing a variable $\hat{\Psi} = \hat{P}/a$ and by choosing the function $a$ adequately (see Appendix~\ref{sec:normal}). We then obtain a wave equation for acoustic and gravito-inertial waves in a
uniformly rotating, centrifugally deformed,
barotropic star as follows:
\beqa
\label{great0}
\begin{aligned}
 \Delta\hat \Psi = - \frac{\omega^2}{c_{\rm s}^2} \hat{\Psi} + \frac{f^2}{\omega^2}\n_{z}^2\hat \Psi + \frac{N_0^2}{\omega^2} \Delta_{\perp} \hat \Psi + i \frac{\|\vec{f}\wedge\vec{g}_0\|}{\omega c_{\rm s}^2} {\cal T} \vec e_{\phi}  \cdot \nab \hat \Psi &\\
+  {\cal C'} \hat \Psi,\quad&
\end{aligned}
\eeqa
where $\vec e_{\phi} $ is the unit vector in the azimuthal direction and ${\cal T}$ is given by Eq.~(\ref{T}).
The first term on the right-hand side (RHS) of \eq{great0} is associated with sound waves whereas the next two terms on the RHS produce gravito-inertial waves. The fourth term on the RHS 
is non-zero only for non-axisymmetric perturbations in rotating stars. It contains 
the latitudinal derivative of the Coriolis parameter $f\cos\theta$, which gives rise
to Rossby and Kelvin waves. 
Such a simple form for the wave equation comes at the cost of very complicated expressions for $a$ and ${\cal C'}$ as functions of $c_{\rm s}, N_0,\vec g_0 , \vec{f}$, and
$\omega$ (see Eqs.~(\ref{eq:a}) and (\ref{eq:const})).

\subsection{The short-wavelength approximation of the perturbation equations}
\label{sec:wkb}

The ray model results 
from the WKB approximation of the wave equation (\ref{great0}).
It consists in looking for 
wave-like solutions
of the form 
\begin{equation} \label{eq:WKB}
\Psi = \Re\lbrace A(\vec x) \exp i[\Phi (\vec x) - \omega t] \rbrace = \Re\lbrace \hat{\Psi}(\vec x) \exp (- i \omega t)\rbrace,
\end{equation}
assuming that the associated wavelength $\lambda_{\rm w} \sim \|\nab \Phi\|^{-1}$ is much shorter
than the typical length scale on which the background medium varies, hereafter denoted $L_{\rm b}$.

The phase $\Phi$ and the amplitude $A$ of $\hat{\Psi}$ 
are expanded into power series of $1/\Lambda$, where $\Lambda=L_{\rm b}/\lambda_{\rm w}$ is assumed to be large, yielding\begin{equation} \label{exp}
\Phi= \Lambda \left( \Phi_0 + \frac{1}{\Lambda} \Phi_1+\hdots\right) \quad\text{and}\quad  A = A_0 + \frac{1}{\Lambda} A_1 +\hdots
\end{equation}
Introducing these expansions into the wave equation \eq{great0} 
and considering only the dominant $\mathcal{O}(\Lambda)$ terms, we obtain
an equation
for $\Phi_0$, the so-called 
eikonal equation. The 
amplitude $A_0$ is determined at the next order.
The eikonal equation obtained with this procedure depends on the frequency range considered.
If $\omega$ is of
the order of $\Lambda$, the dominant terms of the wave equation~\eq{great0} are the left-hand side (LHS) and the first term on the RHS, leading to
an eikonal equation for sound waves $\omega = c_{\rm s} k$, where $\vec{k} = \nab \Phi_0$ and $k=\|\vec{k}\|$.
If $\omega = \mathcal{O}(1)$, the dominant terms are those
that 
contain second derivatives of $\hat{\Psi}$, leading to an eikonal equation
for gravito-inertial waves. 
The fourth term on the RHS of the wave equation~\eqref{great0}
is a first-order azimuthal derivative of $\hat{\Psi}$ and is therefore negligible when $\omega = \mathcal{O}(1)$. Thus, 
Rossby waves are not included in the eikonal equation
if one considers the $\omega = \mathcal{O}(1)$ regime or if one restricts to axisymmetric perturbations.
Before writing down the eikonal equation for gravito-inertial waves, the role of the so-called constant term ${\cal C'} \Psi$ must be clarified.
Indeed, if a wave, acoustic or gravito-inertial, of vanishingly small wavelength approaches the stellar surface, it will cross the surface without being refracted back into the star.
This regime corresponds to the 
limit $\lambda_{\rm w} \rightarrow 0$, where the constant term ${\cal C'} \Psi$ is negligible and the resulting eikonal equation contains no surface  
refraction effect.

We are interested, however, in waves that are refracted back because oscillations modes are the result of wave interferences within the star. 
As known from previous studies on acoustic and gravity waves \citep{AC10}, this back-refraction occurs for outwards-travelling waves that encounter atmospheric layers whose pressure scale heights 
are smaller that the wavelength of the wave. We also know that the pressure scale height strongly decreases as one approaches the surface in such a way that if 
$L_{\rm b}$ is of the order of the pressure scale height in the interior and $H_{\rm s}$ is the
surface pressure scale height, we obtain $H_{\rm s}/L_{\rm b} \ll 1$.
Thus, there exists a wavelength range for which we have at the same time $\lambda_{\rm w}/ L_{\rm b}= 1/\Lambda \ll 1$  inside the star 
and $\lambda_{\rm w}$ of the order of or larger than $H_{\rm s}$.
In the wave equation, this surface refraction effect is accounted for by the constant term ${\cal C'} \Psi$ because ${\cal C'}$ is inversely proportional to the square of the pressure scale height (see Eq.~\eq{cst} below). Thus,
an eikonal equation containing this term is simply obtained by 
assuming 
$\lambda_{\rm w}= \mathcal{O}(H_{\rm s})$ in the expansion given by Eq.~\eq{exp}. This yields
\beqa
\label{eq:eikonal}
k^2 = \frac{f^2}{\omega^2} k_z^2 + \frac{N_0^2}{\omega^2} \left(k_{\perp}^2 + k_{\phi}^2\right) - {\cal C} \qquad \mbox{with} \;\; {\cal C} = {\cal C'} -  \frac{\omega^2}{c_{\rm s}^2},
\eeqa
where $k_z = \vec e_z  \cdot \vec{k}$, $k_{\phi} = \vec e_{\phi}  \cdot \vec{k}$ and $k_{\perp} = \vec e_{\perp}  \cdot \vec{k}$, where the unit vector $\vec e_{\perp} $ is defined as $\vec e_{\perp}  = \vec e_{\phi}  \wedge \vec e_{\parallel} $. 
In Eq.~\eqref{eq:eikonal}, the term in $k_\phi^2$ results from the fact that $\vec\nabla_\perp$ is not just the derivative along $\vec e_\perp$.
It also contains derivatives in $\phi$, and so does $\Delta_\perp$.

Although the full expression of ${\cal C}$ is a very complex function of the equilibrium quantities $c_{\rm s}, N_0,\vec g_0 , \vec f$, and of $\omega$
(see Eq.~\eqref{eq:const}), we only need to estimate ${\cal C}$ where it becomes of the same order of magnitude as the other terms in the dispersion relation, that is near the surface.
The increase of ${\cal C}$ towards the surface is due to the sharp increase of $N_0$ and $1/c_{\rm s}$, while $\vec g_0 $ does not vary much in the stellar envelope.
In a polytropic model of star, we also have $N_0 \propto 1/c_{\rm s}$ near the surface. In order to only retain the dominant term, we look for an expansion of
${\cal C}$ in powers of $c_{\rm s}$ and find that ${\cal C}$ can be written as the ratio between two polynomials of $c_{\rm s}^2$, namely
\beqa
{\cal C} = \frac{C_0 + C_1 c_{\rm s}^2 + \mathcal{O}(c_{\rm s}^4)}{\left(\alpha K + G c_{\rm s}^2\right)^2 c_{\rm s}^4},
\eeqa
where the terms $\alpha=\frac{\Gamma_1\mu}{\mu+1}-1$,
$K=(\vec f\cdot\vec g_0)^2-\omega^2{g_0}^2$, $G=\omega^2(\omega^2-f^2)$ and the terms $C_0$ and $C_1$ are $\mathcal{O}(1)$ at the surface. 
In the super-inertial
\footnote{The terms super-inertial and sub-inertial are also used 
in the geophysical literature but with different meanings as they 
refer to frequencies higher or smaller than $2 \Omega \cos \theta$.}
regime $\omega > f$, $K$ does not vanish. The dominant term of ${\cal C}$ is thus $\frac{C_0}{(\alpha K)^2 c_{\rm s}^4}$ and
we obtain
\beqa
\label{cst}
{\cal C} = \left(1 - \frac{f^2}{\omega^2} \cos^2 \Theta\right) \frac{\Gamma_1^2}{4} \frac{\mu-1}{\mu+1} \frac{g_0^2}{c_{\rm s}^4} + {\cal O} (1/c_{\rm s}^2),
\eeqa
where $\cos \Theta = - \vec f .\vec g_0 / f g_0$.
In the sub-inertial regime $\omega < f$, both $K$ and $C_0$ vanish at the critical angles $\Theta_{\rm c}$ and $\pi -\Theta_{\rm c}$
such that $\cos^2 \Theta_{\rm c} = \omega^2/f^2$. In this regime and in the regions close 
to
the stellar surface and the critical angles, ${\cal C}$ is given by
\beqa
{\cal C} = \frac{C_1(\Theta_{\rm c})}{G^2} \frac{1}{c_{\rm s}^6(\Theta_{\rm c})} + {\cal O} (1/c_{\rm s}^4).
\eeqa
The detail of the asymptotic expression of ${\cal C}$ is presented in Appendices~\ref{sec:const} and \ref{sec:const2}.
In the following, we always use the expression \eq{cst} for ${\cal C}$. 
As we see in the next sections, inspection of the eikonal and ray equations near the critical angle
indicates that rays tend to avoid the surface layers close to the critical angle. This is further confirmed by our computations that do not show rays reaching the surface close 
to the critical angle.

Using \eq{cst}, the eikonal equation can be written as
\beqa
\label{disp0}
\omega^2 = f^2\frac{k_z^2}{k^2+k_{\rm c}^2} + N_0^2 \frac{k_{\perp}^2+k_{\phi}^2}{k^2+k_{\rm c}^2} + f^2\cos^2 \Theta \frac{k_{\rm c}^2}{k^2+k_{\rm c}^2},
\eeqa
with 
\beqa
\label{kc}
k_{\rm c}^2 = \frac{\Gamma_1^2}{4} \frac{\mu-1}{\mu+1} \frac{g_0^2}{c_{\rm s}^4}.
\eeqa

This equation is similar to the local dispersion relations generally used to study gravito-inertial waves in the Earth's atmosphere \citep{FA03} and to construct 
associated ray models \citep{ME95}. 
The present relation is nevertheless
more general in that it takes the centrifugal deformation of the star into account and does not make the traditional approximation.
In the following we shall restrict our study to axisymmetric perturbations, that is to the case $k_{\phi}=0$. 

\subsection{Domains of propagation}
\label{sec:domain}

The eikonal equation (\ref{disp0}) allows us to discuss the domain of propagation of gravito-inertial waves in a more general context
than previous works.
In particular, when compared to the work of \citet{DR00}, the following discussion includes the effects of the centrifugal deformation and, through the $k_{\rm c}$ term, those of the background compressibility.

According to \eq{eq:WKB}, non-evanescent wave solutions are possible only when both components of the wave vector $\vec{k}$ are real numbers. The region of space where
the eikonal equation admits real solutions thus determines the domain where rays can propagate. After projecting $\vec{k}$ onto an orthogonal basis,
the eikonal equation takes the general form of a conic. Using $(\vec e_{\parallel} , \vec e_{\perp} )$ as a basis
and the relations $k^2 = k_{\parallel}^2 + k_{\perp}^2$ and $k_z = k_{\parallel} \cos \Theta - k_{\perp} \sin \Theta$, the eikonal equation (\ref{disp0})
is rewritten as
\begin{equation}
\begin{aligned}
\label{disp1}
(N_0^2 + f^2 \sin^2 \Theta - \omega^2) k_{\perp}^2   -2 f^2 \cos \Theta \sin \Theta k_{\parallel} k_{\perp} \\
- (\omega^2 - f^2\cos^2 \Theta) k_{\parallel}^2 
- (\omega^2 - f^2\cos^2 \Theta) k_{\rm c}^2 =0.
\end{aligned}
\end{equation}

To discuss the constraints on the domain of propagation that can be derived from this relation, it is useful to first consider the
non-rotating case. The eikonal equation then becomes
\beqa
\label{rot0}
\left[N_0^2(r) - \omega^2\right] \frac{L^2}{r^2} - \omega^2 k_r^2 = \omega^2 k_{\rm c}^2(r),
\eeqa
with $k_\perp=k_\theta=\pm L/r$ and $L$ is the norm of the vector $\vec{x}\wedge\vec{k}$, which is equivalent to the angular momentum.
This equation corresponds to the high-sound-speed limit of eikonal equations already derived in the non-rotating case, for example by \citet{G93}.
The condition that $k_r^2$ is positive translates into $\omega^2< N_0^2/(1+r^2k_{\rm c}^2/L^2)$ and this inequality fully specifies the spatial 
limit of
the resonant cavity for a wave of given $\omega$ and $L$. In internal regions where $r^2k_{\rm c}^2/L^2 \ll 1$, it further simplifies 
into $\omega < N_0$, a condition that no longer depends on $L$  or $k_\theta$. 
Closer to the surface, $\omega \ll N_0$ and the condition for propagation is 
$\omega < c_1 S_L$, where $S_L = L c_{\rm s}/r$ is the Lamb frequency and $c_1$ a constant that can be expressed in terms of $\Gamma_1$ and $\mu$.
If this last condition is fullfilled at the surface, the gravity ray escapes the star. 
Otherwise, 
the ray is refracted back into the star and the relation $\omega = c_1 S_L$
determines the outer radius of the domain of
propagation. An important point is that the condition for the back-refraction in the outer layers of the star depends on $L$ or equivalently on $k_\theta$.

Coming back to the rotating case, the condition that $k_{\parallel}$ is real can be expressed as
\beqa
\label{del}
\delta = \Gamma k_{\perp}^2 - (\omega^2 - f^2\cos^2 \Theta)^2 k_{\rm c}^2 \ge 0,
\eeqa
where $\delta$ is the reduced discriminant of the eikonal equation viewed as a quadratic equation for the variable $k_{\parallel}$, and
\beqa
\Gamma = \omega^2 (f^2 -\omega^2) + N_0^2 (\omega^2 - f^2 \cos^2 \Theta). 
\eeqa

The sign of $\Gamma$ determines whether gravito-inertial waves can propagate in the inner region where $k_{\rm c}$ can be neglected.
In contrast, the outer limit of the domain of propagation is determined by the condition $\delta \ge 0$ and,
as expected from the condition of propagation in the non-rotating case, it depends on the value of $k_\perp^2$.
However, contrary to the non-rotating case, $k_\perp^2$ is not known \emph{a priori} because rotation breaks the spherical symmetry and 
the associated conservation of the norm of the angular momentum. 
The consequence is that in
the general case of gravito-inertial waves in a rotating star, we are not able to predict the outer limit of the resonant cavity
from the eikonal equation alone. It is thus necessary to solve the ray dynamics to find out 
whether a ray of a given frequency remains confined within the star and, if this is the case,  the shape of the outer limit of its resonant cavity. 
In the ray dynamics calculation  presented in this paper, all rays stay inside the stars 
because, at the surface of polytropic models, $k_{\rm c} $ is infinite and $\delta$ is thus necessarily negative.
In real stars, $k_{\rm c}$ remains finite at the surface, and this implies that rays with sufficently high values of $k_\perp$ near the surface can propagate outside the stars.
Waves associated with such rays are expected to have very high spatial frequencies at the surface, and are thus unlikely to be visible.
However, they can play a role in the transport of angular momentum out of the stars.

In the following of the section, we thus only discuss the domain of propagation in the internal region where the $k_{\rm c}^2$ term can be neglected, since this
approximation is 
valid from the centre up to       
a certain radius that depends on the given ray.
The condition of propagation $\Gamma \ge 0$ is equivalent to
\beqa
\label{concond}
\omega_{-}^2 < \omega^2 < \omega_{+}^2,
\eeqa
with
\beqa
\omega_{\pm}^2 =\frac{f^2 + N_0^2 \pm \sqrt{\left(f^2+ N_0^2\right)^2 - 4 f^2 N_0^2 \cos^2 \Theta}}{2}.
\eeqa

If
the centrifugal deformation is neglected, that is if we replace $\Theta$ by the colatitude $\theta$,
this condition of propagation is equivalent to that found by \citet{DR00}.
Two limit cases, $f \ll N_0$ and $N_0 \ll f$, are relevant since the former occurs in the bulk of typical moderately rotating stellar radiative zones, 
whereas the latter is verified in 
convective zones or close to the centre of the star where $N_0$ vanishes. If $f \ll N_0$, the conditions \eq{concond} simplify to
\beqa
\label{approx1}
f^2 \cos^2 \Theta  <  \omega^2 < N_0^2 + f^2 \sin^2 \Theta.
\eeqa
On the other hand, if $N_0 \ll f$, the conditions \eq{concond} become
\beqa
\label{approx2}
N_0^2 \cos^2 \Theta <  \omega^2 < f^2 + N_0^2 \sin^2 \Theta,
\eeqa 
where in the limit of vanishing $N_0$, the $\omega < f$ rule for pure inertial waves is recovered.

We now consider the particular case of uniformly rotating $\mu=3$ polytropic models. 
Close to the centre of the star, 
the approximate conditions \eq{approx2} hold because $N_0$ vanishes there. At some distance from the centre, 
the Brunt-V\"{a}is\"{a}l\"{a} frequency
always becomes larger than $f$. Moreover, for moderate rotators, the
approximate condition \eq{approx1} is valid from a small radius up to the surface.
Indeed, as long as $\Omega \le 0.38 \Omega_{\rm K}$ and $r > 0.1 R_{\rm e}$, the ratio $f^2/N_0^2$ is always smaller than $0.10$.

\begin{figure}
\resizebox{\hsize}{!}{\includegraphics{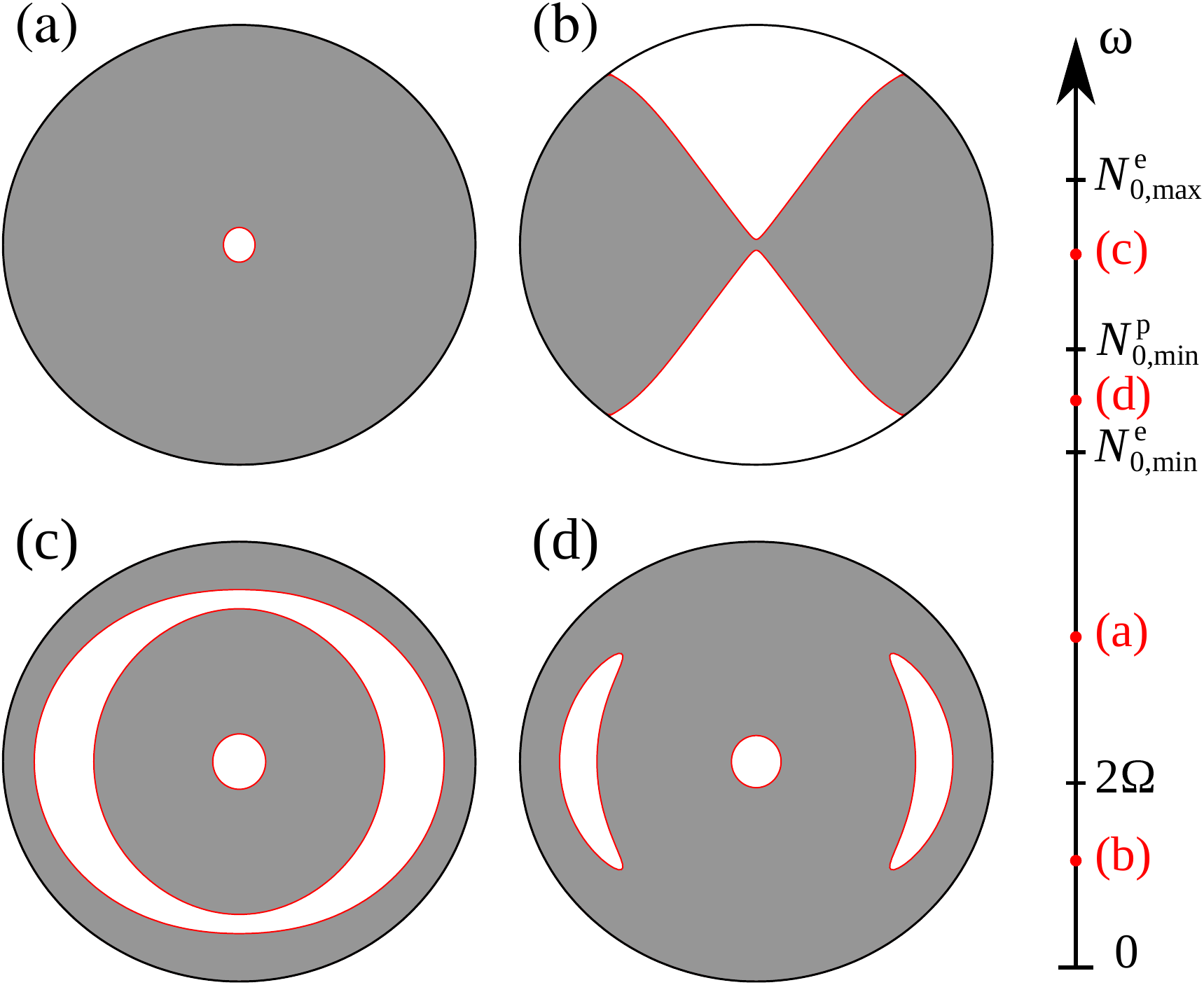}}
\caption{Typical domains of propagation (grey areas) given by the $\Gamma(\omega) \ge 0$ 
condition as a function of the gravito-inertial wave frequency $\omega$ in a $\Omega = 0.38 \Omega_{\rm K}$ polytropic model of star.
In general, the outer limit of the domain, here loosely represented by the stellar surface, cannot be predicted from $\omega$ alone and 
is obtained by solving the ray dynamics.
Panel (a) corresponds to a super-inertial frequency ($\omega>f$) and panel (b) to a sub-inertial frequency ($\omega < f$). The two bottom figures show domains with an intermediate forbidden zone when frequencies are
between the local minimum and maximum of the Brunt-V\"{a}is\"{a}l\"{a} frequency.
Panel (c) corresponds to frequencies in  the $N^{\rm p}_{0,{\rm min}} < \omega < N^{\rm e}_{0,{\rm max}}$ interval
and also exists in non-rotating stars, whereas
panel (d) corresponds to frequencies in the $N^{\rm e}_{0,{\rm min}} < \omega < N^{\rm p}_{0,{\rm min}}$ interval and is due to the centrifugal deformation
of rotating
stars.}
\label{domains}
\end{figure}
Two typical domains of propagation are shown in panels (a) and (b) of Fig.~\ref{domains} in the super-inertial and sub-inertial frequency ranges,
respectively.
The rotation of the stellar model is $\Omega = 0.38 \Omega_{\rm K}$.
For the super-inertial case, the condition of propagation reduces to $\omega^2 < N_0^2 + f^2 \sin^2 \Theta$, which leads to
the domain shown in panel (a), including
the weakly prolate avoided central region. By contrast, in the sub-inertial regime the centre is never forbidden because conditions \eq{approx2} 
are automatically fulfilled for
$\omega < f$. Away from this central region, the conditions of propagation \eq{approx1} reduce to $f^2 \cos^2 \Theta < \omega^2$ for these waves. The resulting domain
of propagation is shown in panel (b). A noticeable effect of the centrifugal deformation is that the limit of the domain is no longer strictly conical 
away from the central region. As the centrifugal deformation of the equipotentials becomes significant towards the surface, the boundary of the domain of propagation 
is bent towards lower latitudes  to
keep $\Theta_{\rm c}=\arccos(\omega/f)$, the angle between the direction of the effective gravity and the rotation axis, constant.

The local minimum of the Brunt-V\"{a}is\"{a}l\"{a} frequency present in the stellar envelope (see Fig.~\ref{Brunt}) induces an additional
forbidden region for frequencies between $N^{\rm e}_{0,{\rm min}}$ and $N^{\rm p}_{0,{\rm max}}$. As illustrated on panel (c) for the $\Omega = 0.38 \Omega_{\rm K}$ model,
the domains of progagation below and above the position of $N^{\rm e}_{0,{\rm min}}$ are disjoint
when $N_{0,{\rm min}} < \omega < N_{0,{\rm max}}$ in non-rotating stars or 
when $N^{\rm p}_{0,{\rm min}} < \omega < N^{\rm e}_{0,{\rm max}}$ in rotating stars. In centrifugally deformed stars, waves with frequencies in the range
$N^{\rm e}_{0,{\rm min}} < \omega < N^{\rm p}_{0,{\rm min}}$ also show a forbidden region, 
but this time it is limited to equatorial regions so that the inner and outer
domains communicate through polar regions. An example is shown in panel (d) for the
$\Omega = 0.38 \Omega_{\rm K}$ model also. We observe
that the outer limit of the forbidden region is oblate whereas the inner limit is prolate.
As can be inferred from the distribution
of $N_0$ shown in Fig.~\ref{Brunt}, this property 
is related 
to
the oblate and prolate
shapes of the iso-contours of 
the Brunt-V\"{a}is\"{a}l\"{a} frequency taken in the $\left[N^{\rm e}_{0,{\rm min}}, N^{\rm p}_{0,{\rm min}}\right]$ interval.
These peculiar resonant cavities appear in a frequency range that increases
with rotation as the ratio $N^{\rm p}_{0,{\rm min}}/N^{\rm e}_{0,{\rm min}}$ goes from $1$ in the non-rotating case 
to $1.70$ at $\Omega = 0.84 \Omega_{\rm K}$.

\section{The ray model}
\label{sec:raymodel}

In this section, the ray model for progressive gravito-inertial waves is derived from the eikonal equation. Its Hamiltonian structure is emphasised in Sect.~\ref{sec:Hamilton} as
it enables us to investigate ray paths 
using tools and concepts developed for Hamiltonian dynamical systems. The geometrical structure of the phase space in particular is the key to
understand the nature of the dynamics. Despite 
its high dimensionality, the phase space can be explored numerically through cuts known as Poincaré surfaces of section (hereafter PSS). 
The numerical method and
PSS are presented 
in Sects.~\ref{sec:numerics} and \ref{sec:pss}, respectively.

\subsection{The Hamiltonian equations describing the ray path and the wave vector evolution along the path}
\label{sec:Hamilton}

The eikonal equation \eq{disp0} is a partial differential equation (PDE) for the phase function $\Phi_0(\vec{x})$. Instead of solving this equation as a PDE,  
the gravito-inertial ray model consists in searching for solutions along some path $\vec{x}(t)$ called the ray path. One then has
to solve the coupled differential equations that govern the ray path and the evolution of $\vec{k}$ along the path.

It has been shown that this problem can be written under very general circumstances in a Hamiltonian form \citep{L78, O02}.
In the present case, the Hamiltonian is obtained by writing the eikonal equation in the form $\omega = H(\vec{x}, \vec{k})$
and the ray path is defined by the group velocity. For a given coordinate system $[x_i]$, the
Hamiltonian equations of the ray dynamics are \citep{L11}
\beqa
\label{Ham0}
\frac{{\rm d} x_i}{{\rm d} t} & = & \frac{\partial H}{\partial k_i}, \\
\frac{{\rm d} k_i}{{\rm d} t} & = & -\frac{\partial H}{\partial x_i},
\label{Ham0bis}
\eeqa
where the $k_i = \frac{\partial \Phi_0}{\partial x_i}$
are the covariant components of $\vec{k}$ on the natural basis 
$\vec{e}_i = \frac{\partial \vec{x}}{\partial x^i}$ associated with the coordinate system $[x_i]$.
In Appendix~\ref{sec:spherical}, these equations are written with spherical coordinates and the $\vec{k}$ 
components on the usual unit vectors $\vec{e}_r$ and $\vec{e}_\theta$.  

A classical property of gravito-inertial waves is the orthogonality of the group velocity $\vec v_{\rm g} ={\rm d}\vec{x}/{\rm d}t$ and the phase velocity $\vec v_{\rm p}  = \omega \vec{k} /k^2$.
Here, this property is valid in the interior but broken when $k_{\rm c}$ is not negligible. This is apparent in the following formula:
\begin{equation} \label{Perp}
\vec v_{\rm g}  \cdot \vec v_{\rm p}  = \frac{(\omega^2 - f^2 \cos^2 \Theta) k_{\rm c}^2}{k^2 ( k^2 + k_{\rm c}^2)},
\end{equation}
whose derivation is detailed in Appendix~\ref{sec:surface}.

\begin{figure}
\resizebox{\hsize}{!}{\includegraphics{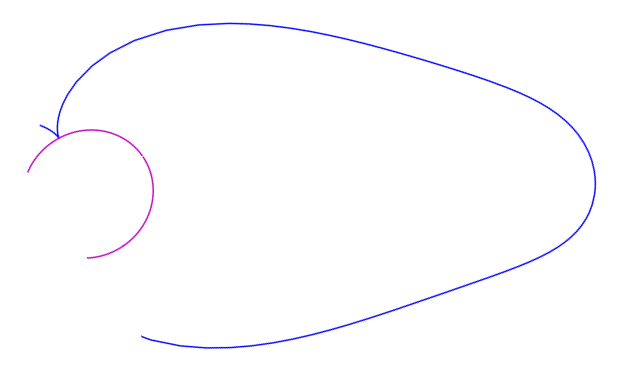}}
\caption{Gravito-inertial ray showing the two types of turning points, a cusp point on the $\Gamma=0$ boundary (in magenta) and a smooth back-refraction in the surface layers
where the decreasing density scale height becomes of the order of the wavelength in that direction.}
\label{rebonds}
\end{figure}

Another important characteristic is the behaviour of the rays whenever they reach the limits of the domain of propagation.
This can be discussed from the relations
\beqa
\label{Retour}
\frac{{\rm d} \vec{x}}{{\rm d} t} \cdot \vec e_{\parallel}  & = & \pm \frac{\sqrt{\delta}}{\omega ( k^2 + k_{\rm c}^2)}, \\
\frac{{\rm d} \vec{x}}{{\rm d} t} \cdot \vec e_{\perp}  & = & \frac{\Gamma k_{\perp} \pm f^2 \cos \Theta \sin \Theta \sqrt{\delta}}{\omega (\omega^2 - f^2 \cos^2 \Theta) ( k^2 + k_{\rm c}^2)},
\label{Retourbis}
\eeqa
also derived in Appendix~\ref{sec:surface}.
Equation~\eq{Retour} shows that the component of the group velocity parallel to $\vec e_{\parallel} $ 
vanishes at the limits of the domain of propagation, that is when $\delta = 0$.
If this occurs when $k_{\rm c}$ is negligible, Eq.~\eq{del} finds that $\Gamma$ also vanishes there. Thus, according
to Eq.~\eq{Retourbis}, both components of the group velocity 
vanish at the limit of the domain of propagation and the turning point is an ordinary cusp point.
However, if $\delta = 0$ is reached closer to the stellar surface, where $k_{\rm c}$ cannot be neglected, Eqs.~\eq{del} 
and~\eq{Retourbis} imply that $\Gamma$ and ${\rm d} \vec{x}/{\rm d} t \cdot \vec e_{\perp} $ 
do not vanish. The
trajectory is then locally parallel to $\vec e_\perp $, that is tangent to an equipotential, before it is redirected towards the stellar interior.
Figure~\ref{rebonds} illustrates the two types of turning points for a given trajectory in the super-inertial regime.
The capacity of the numerical method to accurately compute the trajectories near cusp points has been successfully tested against an analytical solution (see Appendix~\ref{sec:test}).

\subsection{Numerical method for the ray dynamics}
\label{sec:numerics}

The gravito-inertial ray dynamics has been investigated by integrating numerically Eqs.~\eq{Ham1}--\eq{Ham4} using
a 5th-order Runge-Kutta method. The step size of the integration is chosen automatically
to keep the local error estimate smaller than a given tolerance.
The background polytropic model of star is solved numerically with an iterative scheme,
which is described in \citet{RC05}.

We checked that the PSS shown in Sect.~\ref{sec:dynamics} are not significantly modified by decreasing the maximum allowed local error of the Runge-Kutta scheme.
We also checked the influence of the resolution of the background polytropic stellar model.
Finally, the conservation of the Hamiltonian is used as an independent accuracy test
of the
computations.

\subsection{Phase-space visualisation: the Poincar\'e surface of section}
\label{sec:pss}

The gravito-inertial ray dynamics is governed by a Hamiltonian with two degrees of freedom, where the phase space is four-dimensional.
The fact that $\omega = H$ is time independent is equivalent to energy conservation.
It implies that phase-space trajectories actually stay on a
three-dimensional surface. A PSS is the intersection of all phase-space trajectories with a given three-dimensional surface,
defined for example by fixing one coordinate. A PSS at a given frequency
is therefore a two-dimensional surface.

Different choices are possible for the PSS, but to provide a complete view of phase space, it must be intersected by
most phase-space trajectories. Here we used a PSS defined by fixing the colatitude $\theta$ to $\pi/2$, which corresponds to half of the equatorial plane.
Furthermore, to ensure
that two trajectories do not intersect on the same point on the PSS, the intersection with $\theta=\pi/2$ is taken into account only when
trajectories cross it from one particular side (here from the $\theta< \pi/2$ side).
For a given frequency $\omega$, the PSS is computed by following many different
trajectories over long time intervals. Then to scan the phase space at a given rotation, the PSS is computed for
frequencies spanning the $\left[0, N^{\rm p}_{0,{\rm max}}\right]$ interval.
For non-integrable systems, the phase-space structure is complex and the features observed in a PSS are expected to depend on the resolution (in $k_r$, $r$ or $\omega$)
by which the dynamics is investigated. However, modes occupy a finite phase-space volume because Fourier analysis shows that their wave-vector localisation is inversely proportional to
their spatial localisation \citep[see][]{LG2}. Investigating phase space with a finite resolution is therefore sufficient to infer mode properties.

\section{Gravito-inertial ray dynamics in rotating stars}
\label{sec:dynamics}

In this section, we study the evolution of the gravito-inertial ray dynamics with rotation. 
We first describe the case of pure gravity rays in a non-rotating polytropic model (Sect.~\ref{sec:nonrot}) and then consider gravito-inertial rays under the traditional approximation (Sect.~\ref{sec:tradi}). 
These are two simple integrable
systems that 
 serve as a reference to study the general case of gravito-inertial rays 
in a rotating star (Sect.~\ref{sec:general}).

\subsection{The non-rotating case $\Omega=0$}
\label{sec:nonrot}

In the non-rotating case, the eikonal equation simplifies into Eq.~\eq{rot0}, where $L$ is the second invariant
that makes the Hamiltonian system with two degrees of freedom
integrable.
The phase space of integrable systems is made of nested invariant tori. These structures are said to be invariant 
because any trajectory that starts on one of the structures stays on it; they are called tori because they have
the topology of a two-dimensional torus.
Any torus is specified by a value of the doublet ($\omega$, $L$) and its intersection
with the $\theta=\pi/2$ hypersurface is the following $f_{L^2,\omega}(k_r, r)=0$ curve:
\beqa
k_r^2 = \left[\frac{N_0^2(r)}{\omega^2} - 1\right] \frac{L^2}{r^2} - k_{\rm c}^2(r),
\label{curve}
\eeqa
trivially derived from the eikonal equation ~\eq{rot0}.

As for any integrable system, two types of invariant tori exist.
Irrational tori, on which any ray covers the whole torus, or equivalently fills the $f_{L^2,\omega}(k_r,r)=0$ curve on the PSS, and rational (or resonant) tori, on which
any ray closes on itself before covering the torus. Rays of resonant tori are thus periodic orbits that imprint the PSS on a finite number of points.

\begin{figure}
\resizebox{\hsize}{!}{\includegraphics{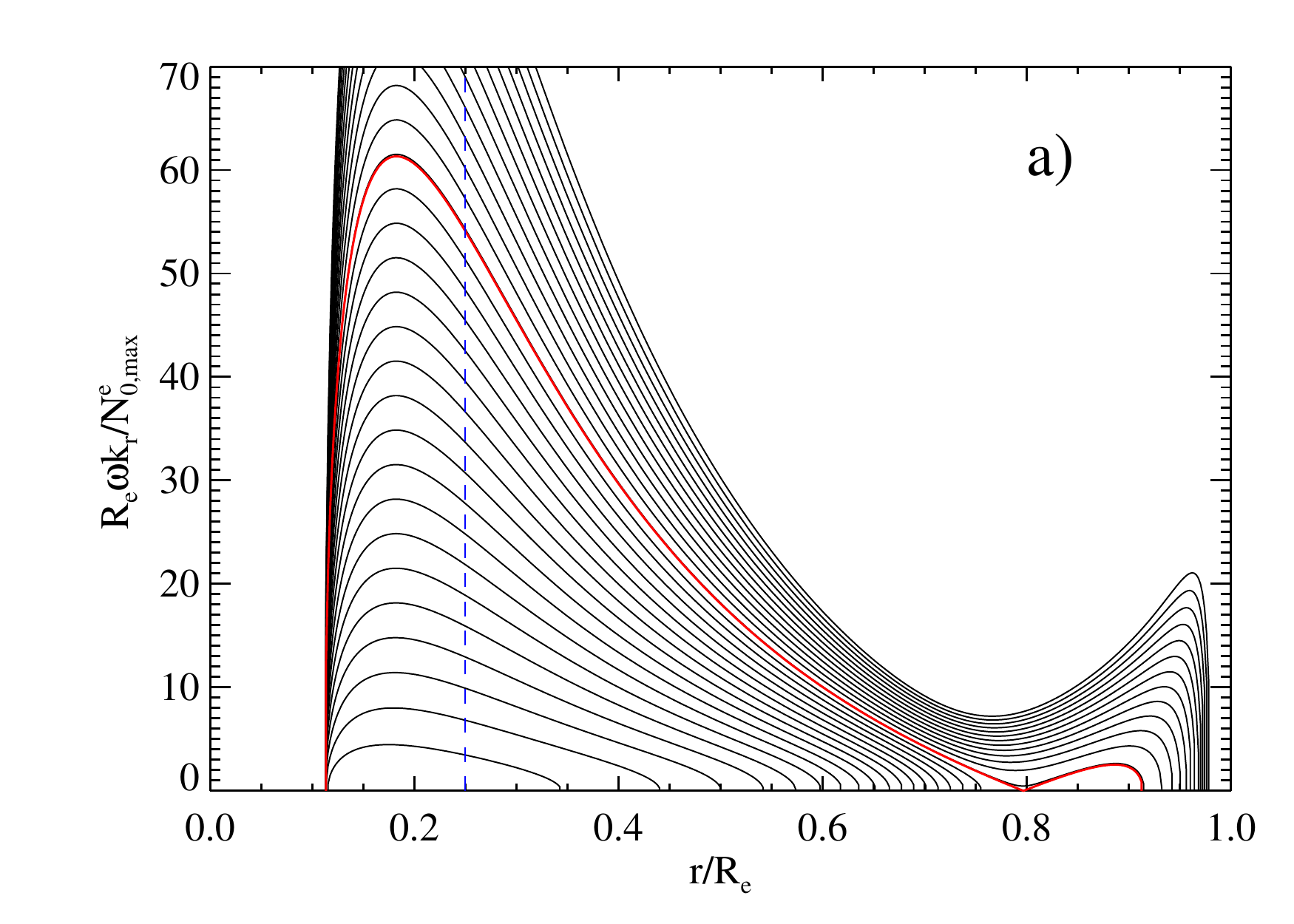}}
\resizebox{\hsize}{!}{\includegraphics{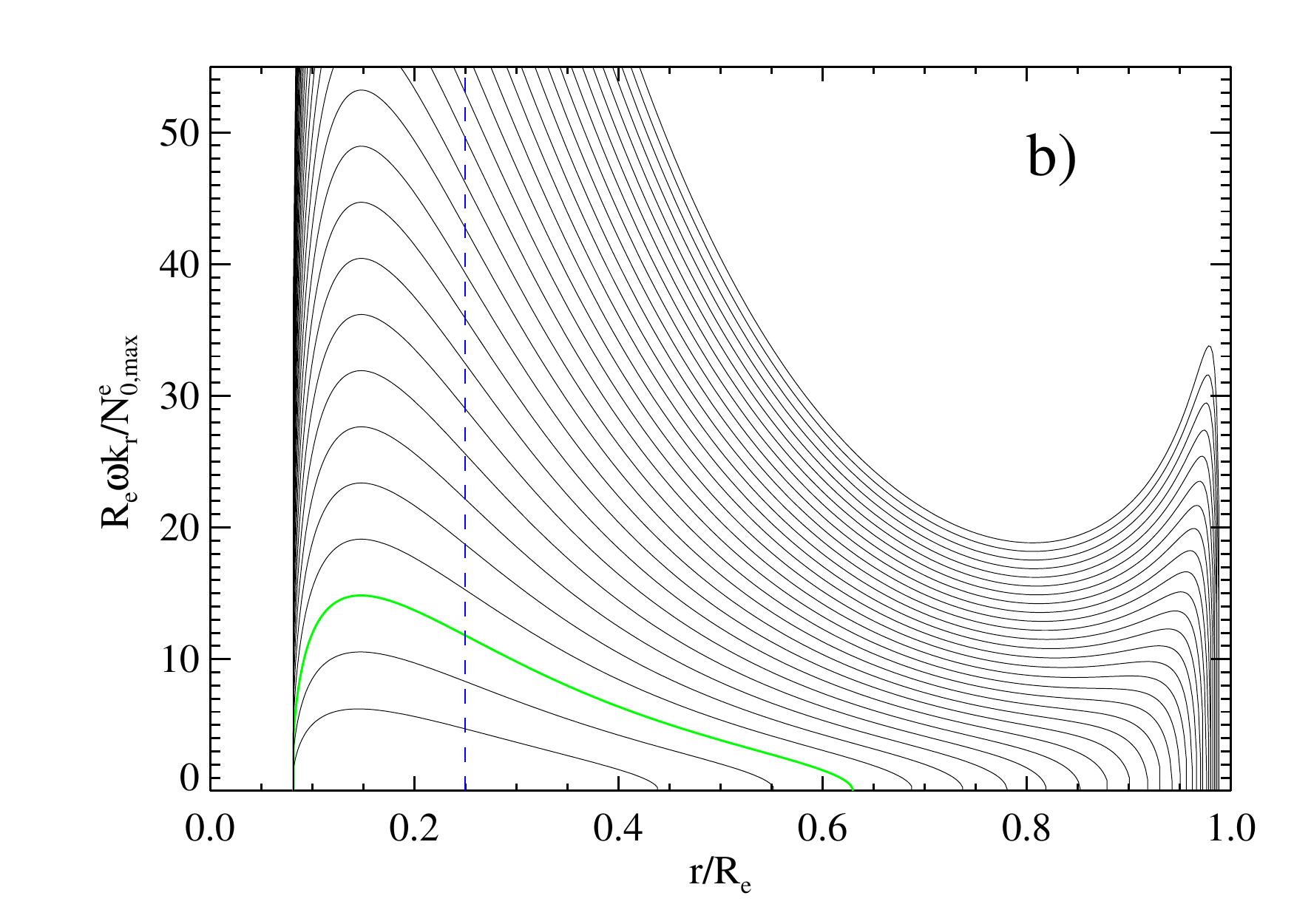}}
\resizebox{\hsize}{!}{\includegraphics{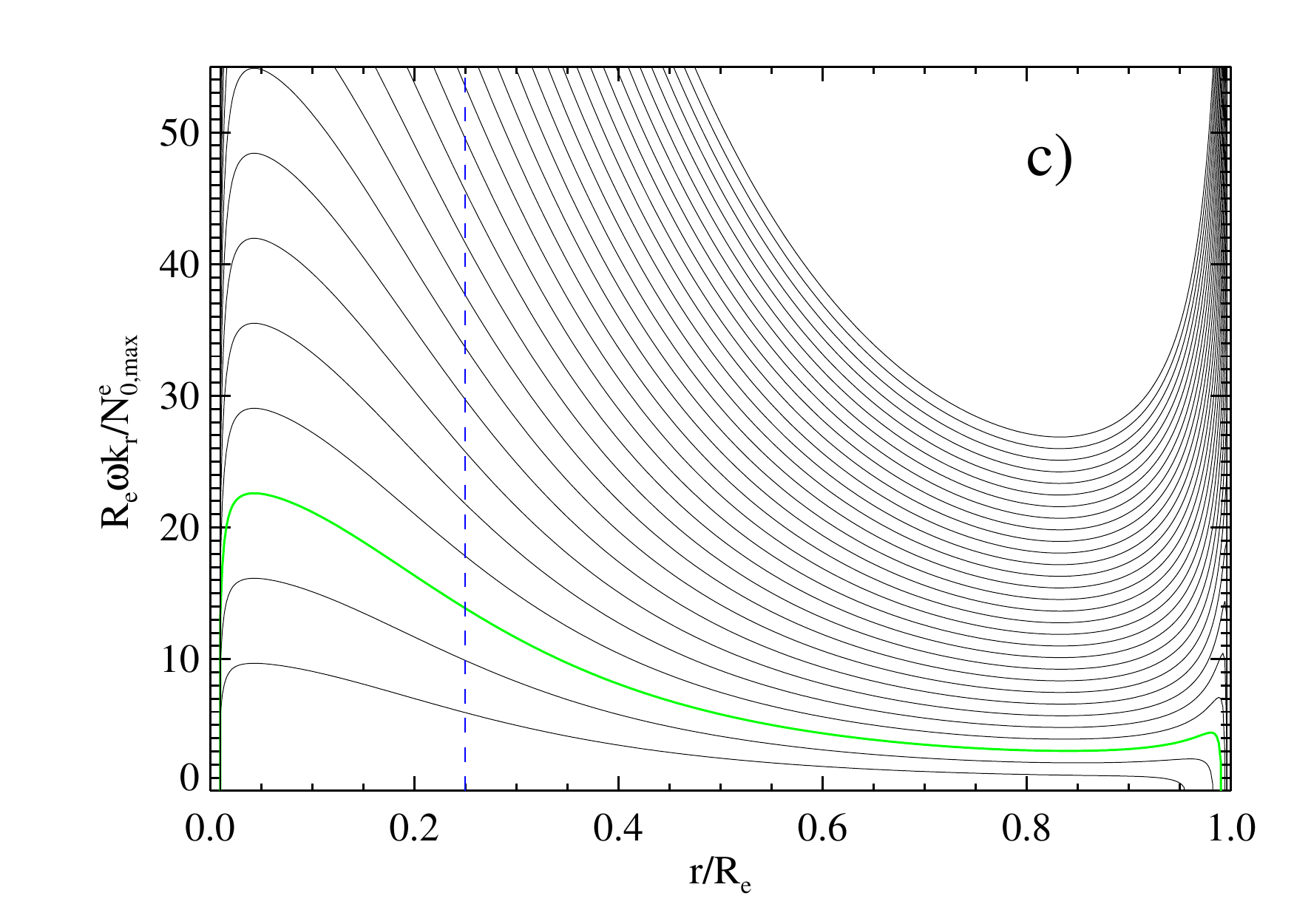}}
\caption{From panel (a) to panel (c), three PSS at $\omega/N_{0,{\rm max}}=[0.66676, 0.50200, 0.05522]$
in a non-rotating polytropic model of star.
The $f_{L^2,\omega}(k_r,r)=0$ curves shown are the imprint on the PSS of the invariant tori
of the integrable gravity ray dynamics. They are drawn here for $L=1/2+\ell$ with $\ell$ taking integer
values 
from $1$ to $30$ corresponding to degrees of spherical harmonics. The value of $L$ of a given tori can be easily estimated from 
the intersection with the blue vertical dashed line, since the ordinate of the intersection point
is approximately equal to $4L$.
The EBK quantisation enables us to specify the tori where
positive interferences of rays produce modes. For example,
the $(n,\ell) = (3,3)$
gravity mode on panel (b) and the $(n,\ell) = (50,3)$ gravity mode on panel (c) are constructed on the tori highlighted in green.
The red curve on panel (a) is a separatrix also called a homoclinic orbit
because it connects the hyperbolic point at $(k_r,r/R)=(0,0.7968)$ to itself.
As explained in Sect.~\ref{cha}, it is at the origin
of a large chaotic phase-space region observed in the rotating case.}
\label{SdPsep}
\end{figure}
Figure~\ref{SdPsep} presents three PSS computed at three different frequencies. For integrable systems, modes
can be related to the dynamics through the Einstein-Brillouin-Keller (EBK) quantisation procedure that has been applied to non-rotating 
spherical stars by \citet{G86} \citep[see also][]{G93,LG2}.
Quantisation conditions
actually select tori that support modes in the sense that rays belonging to these tori interfere positively to construct standing waves, i.e. modes. 
Accordingly, the degree of the spherical harmonic of the mode, $\ell$, is related to the invariant $L$  by the relation $L=\ell+1/2$. We therefore
chose to 
represent in Fig.~\ref{SdPsep} 
tori with $\ell$ between $1$ and $30$. Two of the three PSS have been computed for frequencies corresponding to gravity modes: the $(n,\ell) = (3,3)$ 
and the $(n,\ell) = (50,3)$ modes.
On these PSS, the torus associated with the mode is highlighted in green.
These examples are useful to gain insight into the relation between
the position on the PSS and the modes. In the same spirit,
the $k_r$ normalisation has been chosen in such a way that the value of $L$ of a given $k_r = f_{L^2,\omega}(r)$ curve can be easily retrieved.
Indeed, from the eikonal equation ~\eq{rot0} and provided that
$\omega \ll N_{0,{\rm max}}$, we obtain $R \omega k_r/N_{0,{\rm max}} \simeq L R/r_{\rm max}$, where $r_{\rm max}=0.28 R$ is the radius
of the inner maximum of $N_0$. Thus, the ordinate of the intersection of the $k_r = f_{L^2,\omega}(r)$ curve with the $r=0.25 R$ vertical line is approximatively equal to $4 L$.

\subsection{The traditional approximation}
\label{sec:tradi}

The so-called traditional approximation \citep{E60, LS87} consists in neglecting the contribution of the 
latitudinal component of the rotation vector 
in the Coriolis force.
The solutions of the perturbation equation are then separable into the coordinates $\theta$ and $r$.
This
approximation is justified if motions
are predominantly horizontal, the effect of the Coriolis force in the radial direction is negligible, and the centrifugal deformation is neglected.

Assuming a spherically symmetric stellar model (thus $\vec{e}_{\parallel}=\vec{e}_r$ and $\vec{e}_{\perp}=\vec{e}_{\theta}$) and neglecting the $f \sin\theta$ terms, 
the eikonal equation \eq{disp1} simplifies into
\beqa
\label{dispt}
\left[N_0^2(r) - \omega^2\right] k_\theta^2 - (\omega^2 - f^2 \cos^2 \theta ) \left[k_r^2 + k_{\rm c}^2(r)\right] =0.
\eeqa
From this equation and the associated dynamical equations, we found a second invariant, namely
\beqa
\label{Trad1}
\lambda = \frac{\omega^2 r^2 k_\theta^2}{\omega^2 - f^2 \cos^2 \theta} = \frac{\omega^2 r^2 (k_r^2 + k_{\rm c}^2)}{N_0^2 -\omega^2},
\eeqa
which implies that the associated ray dynamics is integrable.
Integrability is expected since the perturbation equation in the traditional approximation is separable.
Furthermore, neglecting the $k_{\rm c}$ term, the condition of propagation derived from \eq{dispt} is
\beqa
(N_0^2 - \omega^2)(\omega^2 - f^2 \cos^2 \theta) > 0.
\eeqa  
For super-inertial waves, the domain of propagation is simply given by $\omega < N_0$. This frequency range is more restricted than that given by $\Gamma>0$,
but both are qualitatively similar and even identical in the limit 
$f/N_0  \rightarrow 0$  if the centrifugal deformation is neglected (see Eq.~\ref{approx1}). 
For sub-inertial waves, the domain of propagation of the traditional approximation 
consists of two unconnected domains above and below the radius where $\omega=N_0$. Above this radius, the equatorial region defined
by $f \cos \theta < \omega$ and below, the polar cone defined by $\omega < f \cos \theta$.
Thus, the propagation of sub-inertial waves from the 
$N_0 > f$ layers towards the central region  $N_0 < f$ is not allowed by the traditional approximation \citep[for a graphical account of these 
differences in the domain of propagation, see Fig.~11 of][]{GZ08}. This is a clear limitation of
the traditional approximation
since, as we have seen from the discussion on the domain of propagation, sub-inertial waves are indeed allowed to go through the 
centre of the star (and more generally through regions where $N_0 < f$). 
Near the surface, the condition is $\omega < c_1 S_{\!\!\!\sqrt{\lambda}}$ with $S_{\!\!\!\sqrt{\lambda}} = \sqrt{\lambda} c_{\rm s}/r$, which is a condition
formally similar to the non-rotating case, with $\lambda$ playing the role of $L^2$.

\begin{figure}
\resizebox{\hsize}{!}{\includegraphics{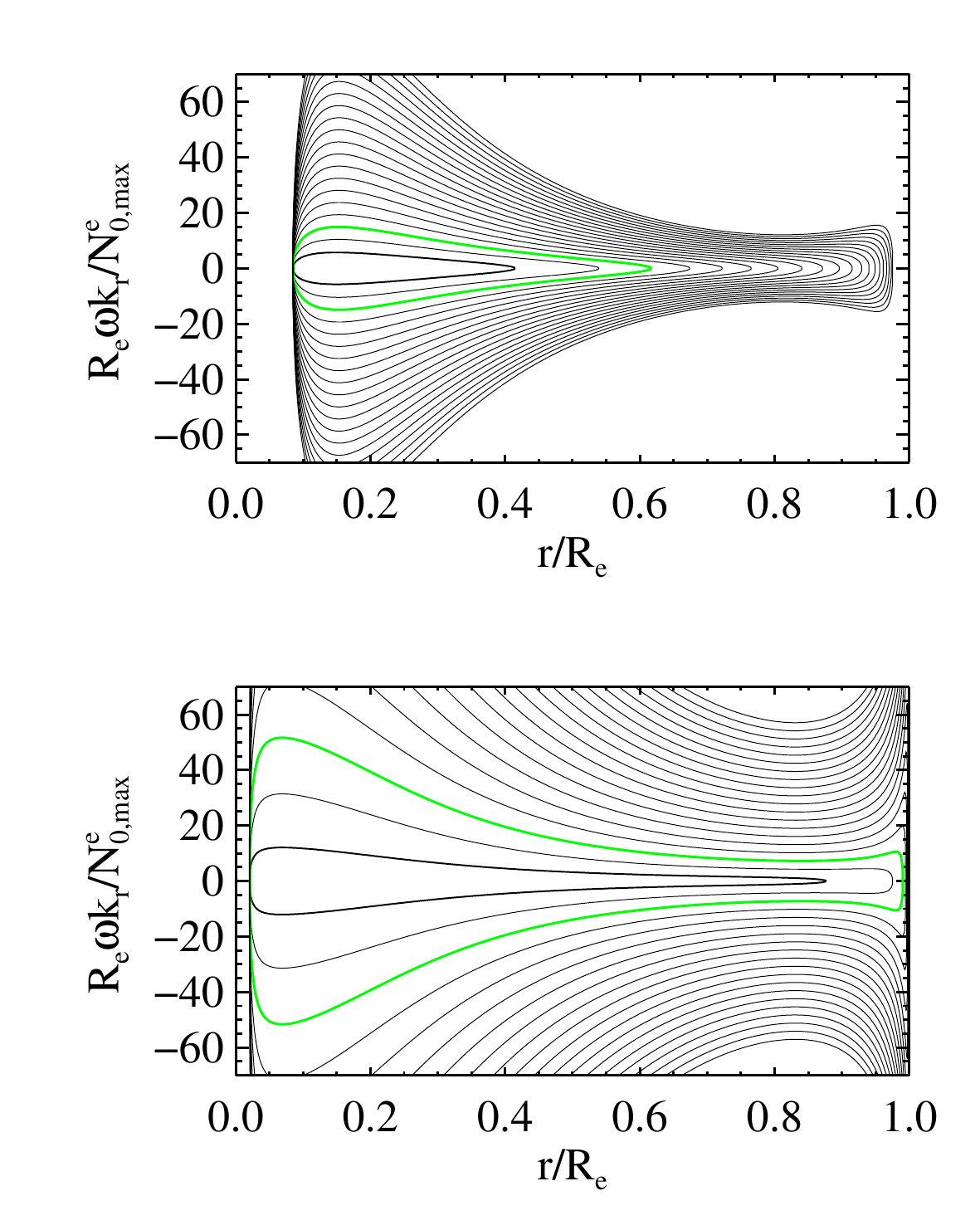}}
\caption{Two PSS at $\omega/f=[2.21,0.567]$
of the ray dynamics in the traditional approximation
in a spherical star rotating at $\Omega = 0.38 \Omega_{\rm K}$.
The $f_{\lambda,\omega}(k_r,r)=0$ curves shown are the imprint on the PSS of the invariant tori
of the integrable dynamics. They are drawn for the quantised values of $\lambda(\ell,\omega/f)$, where $\ell$ is the degree
of the spherical harmonics of the corresponding mode at $\Omega=0$ and is varied from $1$ to $30$.
The increase of the gap between the $f_{\lambda,\omega}(k_r,r)=0$ curves as $\omega$ decreases is due to
the increase of $\lambda$ with $f/\omega$.
Tori in green correspond to gravito-inertial modes labelled by their quantum numbers at zero rotation, that is the $(n,\ell) = (3,3)$
mode (top) and the $(n,\ell) = (50,3)$ mode (bottom).}
\label{SdPTrad}
\end{figure}

The second expression in \eq{Trad1} yields the imprint of the invariant tori on the PSS.
We observe that it has the form $f_{\lambda,\omega}(k_r, r)=0$ that is the same as in the non-rotating case using $\lambda$ instead of $L^2$.
This relation between $\lambda$ and $L^2$ is the ray-dynamics version of the known relation between the separation constant of the perturbation equation 
in the traditional approximation and 
the product $\ell(\ell +1)$ \citep{BD13}.
The PSS of the traditional approximation has thus the same structure as the non-rotating PSS.
The effect of rotation comes from the increase of $\lambda$ with $f/\omega$, which produces
the dilation effect observed between the two PSS shown in Fig.~\ref{SdPTrad}. 
As for the non-rotating case, gravito-inertial modes computed under the traditional approximation can be associated with specific tori.
 This is the case in Fig.~\ref{SdPTrad}, where  two gravito-inertial modes labelled by their quantum numbers at zero rotation, namely $(n,\ell) = (3,3)$
and $(n,\ell) = (50,3)$, are represented in green on the PSS.

\subsection{Phase-space structure of rotating stars}
\label{sec:general}

\begin{figure*}
\centering
   \includegraphics[width=17cm]{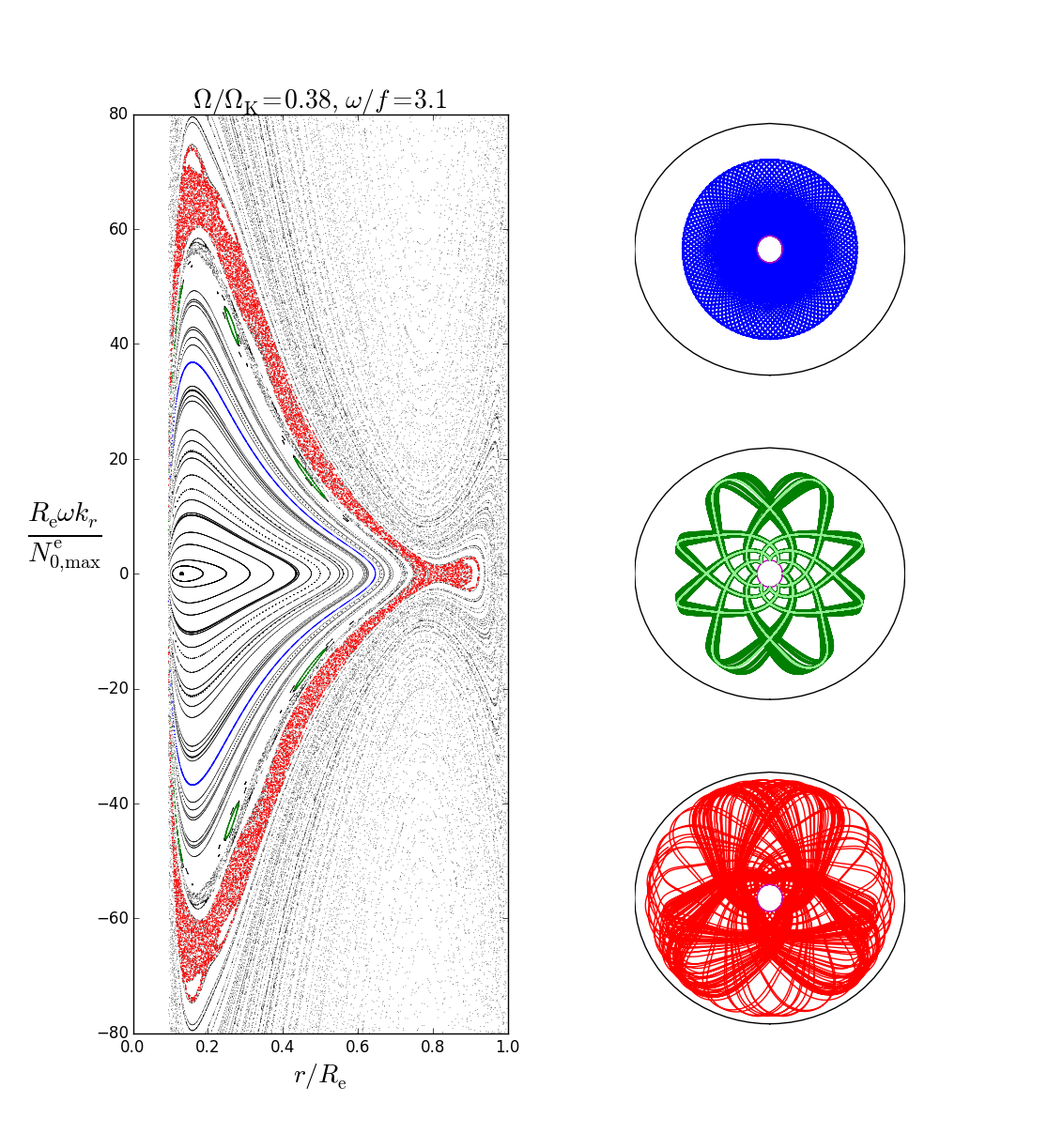}
\caption{PSS computed at  $\omega/f=3.1$ for a rotation $\Omega/\Omega_{\rm K} = 0.38$ and four gravito-inertial rays in the physical space belonging to the three different types of phase-space structures:
a surviving KAM torus (blue), a period-5 island chain (green/light green), and a chaotic zone (red). The light green ray on the middle panel is the stable periodic orbit at
the centre of the period-5 island chain. The period of an island chain is defined as the number of corresponding structures on the PSS (see Sect.~\ref{sec:island}). These rays cross the PSS on points of corresponding colours. On all figures showing rays in the physical space, the limits of the domain of propagation
given by $\Gamma(\omega)=0$ are drawn in magenta.}
\label{SdP31}
\end{figure*}

As rotation increases the gravito-inertial ray dynamics is no longer integrable and undergoes a transition towards 
a mixed state with coexistence of chaotic regions and invariant tori. 
This is apparent on Fig.~\ref{SdP31}, which shows the phase-space structure at $\Omega = 0.38 \Omega_{\rm K}$ and $\omega/f=3.1$ together with
a selection of rays in the
physical space.
As compared to the non-rotating PSS, which only exhibits $f_{L^2,\omega}(k_r, r)=0$ one-dimensional curves, we now observe three different types of structures.
First of all, there are one-dimensional curves similar in shape to the non-rotating curves. The blue ray provides an example of ray associated with this type of tori.
A second and new type of phase-space structures forms around stable periodic orbits and is called
island chains.
An orbit is said to be stable when trajectories induced by small perturbations of the orbit remain close to it.
The green ray highlights one of the invariant tori formed around a central periodic orbit represented by the light green ray  shown in the physical space.
Finally, the third type of structure is the chaotic region, which can be easily identified on the PSS by the fact that the associated rays do no stay on a one-dimensional curve but 
fill a two-dimensional surface instead. The chaotic zone of Fig.~\ref{SdP31} is highlighted by the imprint of a ray also shown in the physical space. 

The apparition of these new structures in the gravito-inertial dynamics is typical of the KAM-type (in reference to the Kolmogorov-Arnold-Moser theorem) transition of Hamiltonian systems from integrability towards chaos \citep{O02}.
Accordingly, in systems that experience infinitesimal departures from integrability, all the resonant tori of the integrable system
are destroyed and replaced by island chains around stable periodic orbits
and small chaotic regions developing around unstable periodic orbits. Meanwhile, irrational tori are continuously perturbed but not destroyed.
The evolution towards larger values
of the parameter controlling the departure from integrability has
been studied in many cases. The common phenomenology emerging from these studies is that, when the control parameter increases,
the invariant tori, either the surviving irrational tori or the island chains, are progressively destroyed while chaotic regions occupy larger phase-space volumes.
As demonstrated in the simple case of the kicked rotor \citep{O02}, for each irrational torus there is a critical value of the control parameter 
above which the torus is destroyed and these critical values
serve to quantify the progression towards chaos.

\begin{figure}
\resizebox{\hsize}{!}{\includegraphics{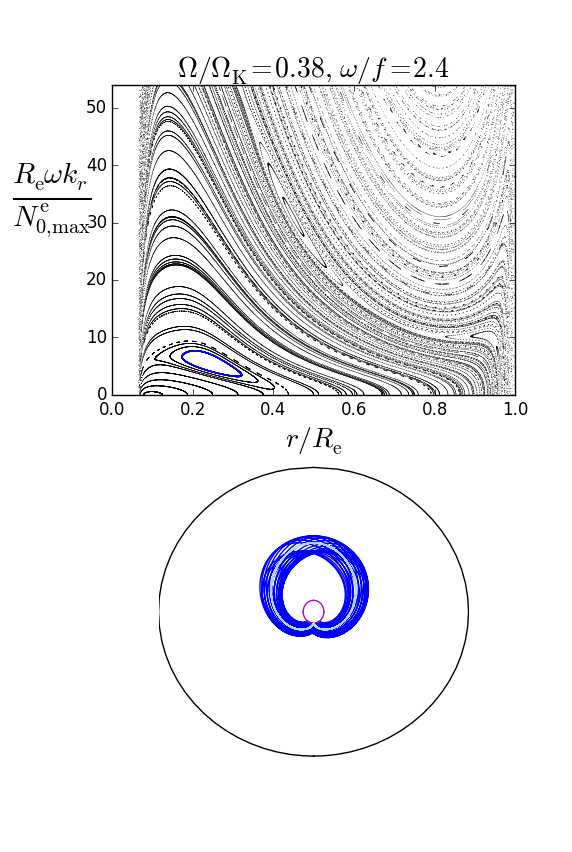}}
\caption{PSS computed at  $\omega/f=2.4$ for a rotation $\Omega/\Omega_{\rm K} = 0.38$ and two gravito-inertial rays belonging to a period-1 island chain including
the central stable periodic orbit (light blue).}
\label{SdP24}
\end{figure}

\begin{figure}
\resizebox{\hsize}{!}{\includegraphics{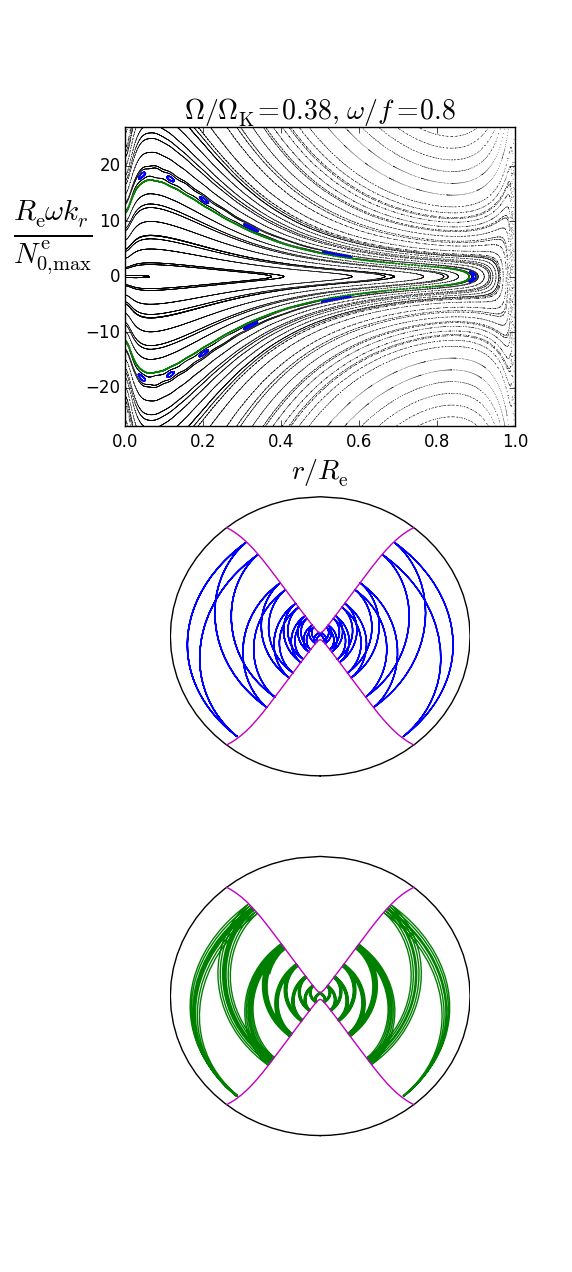}}
\caption{PSS computed at  $\omega/f=0.8$ for a rotation $\Omega/\Omega_{\rm K} = 0.38$ and two gravito-inertial rays  belonging to a period-11 island chain (blue)
and to a surviving KAM torus (green), respectively.} 
\label{SdP08}
\end{figure}

\begin{figure*}
\resizebox{\hsize}{!}{\includegraphics{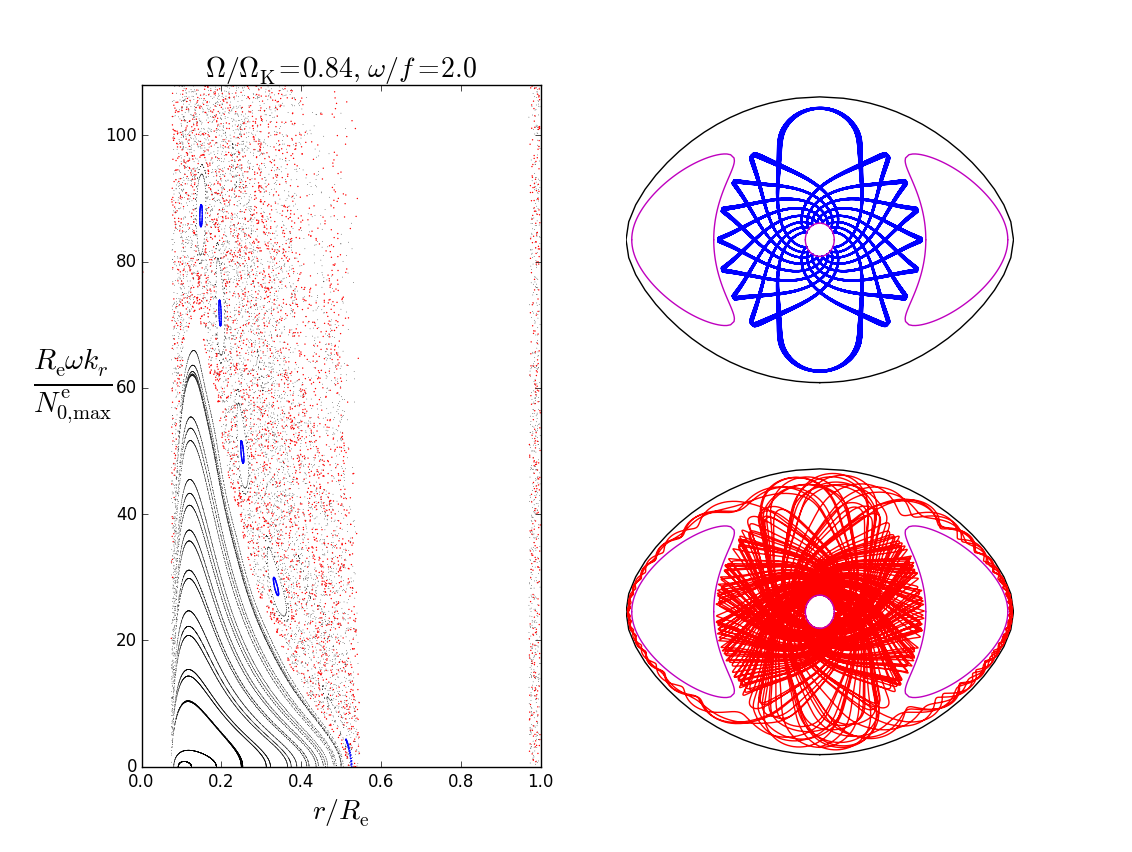}}
\caption{PSS computed at  $\omega/f=2.0$ for a rotation $\Omega/\Omega_{\rm K} = 0.84$ and two gravito-inertial rays  belonging to a period-9 island chain (blue)
and to a chaotic zone (red), respectively.}
\label{SdP20}
\end{figure*}

\begin{figure}
\resizebox{\hsize}{!}{\includegraphics{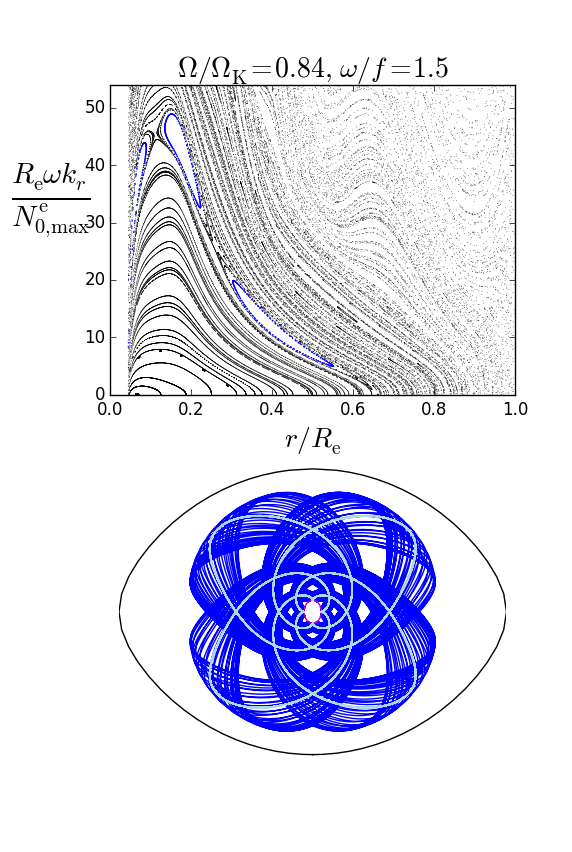}}
\caption{PSS computed at  $\omega/f=1.5$ for a rotation $\Omega/\Omega_{\rm K} = 0.84$ and two gravito-inertial rays belonging to a period-5 island chain including
the central stable periodic orbit (light blue).}
\label{SdP15}
\end{figure}

\begin{figure}
\resizebox{\hsize}{!}{\includegraphics{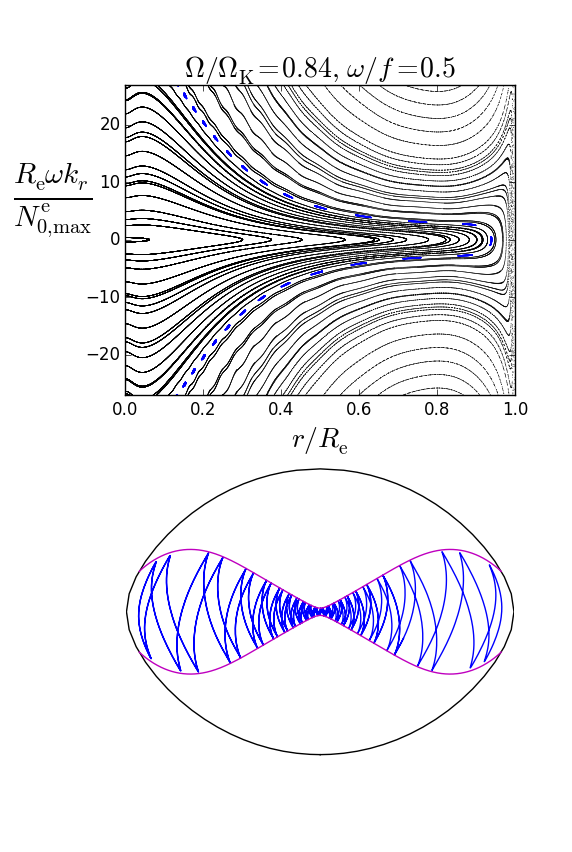}}
\caption{PSS computed at  $\omega/f=0.5$ for a rotation $\Omega/\Omega_{\rm K} = 0.84$ and a gravito-inertial ray belonging to an island chain of large ($> 26$) period.}
\label{SdP05}
\end{figure}

The present  gravito-inertial ray dynamics clearly undergoes a KAM-type transition towards a mixed state including 
surviving irrational tori, island chains, and chaotic regions. Nevertheless, we find that the actual departure from integrability strongly 
depends on the frequency $\omega$. Indeed, the surface occupied by the island chains and  chaotic regions strongly decreases towards
low frequencies. This is obvious when comparing the phase-space structure of a $\Omega = 0.38 \Omega_{\rm K}$ star at three frequencies, $\omega/f= \{3.1,2.4,0.8\}$,
as shown in Figs.~\ref{SdP31},~\ref{SdP24}, and~\ref{SdP08}, respectively. 
This observation does not mean that island chains and chaotic regions are 
absent below some frequency but rather that they are too small to be visible when the PSS are shown on large $k_r$ and $r$ scales.
The same phenomenology also holds at higher rotation as shown by the PSS of a $\Omega = 0.84 \Omega_{\rm K}$ star computed for three decreasing frequencies, 
$\omega/f= \{2, 1.5, 0.5\}$
and shown  in Figs.~\ref{SdP20},~\ref{SdP15},~and \ref{SdP05}, respectively.
We therefore conclude that the departure from integrability as measured by the presence of large island chains and chaotic regions strongly diminishes in
the low part of the $\left[0,N^{\rm p}_{0,{\rm max}}\right]$ interval and is weakly dependent on the centrifugal deformation.

Overall we may say that the tori that results from the continuous evolution of the non-rotating tori still shape 
the large-scale organisation of the phase space even at large
rotation rates.
This is in contrast with the axisymmetric acoustic ray dynamics where, as rotation increases, 
the phase space is progressively dominated by large islands
and a large chaotic region,
while irrational tori only survive for large values of $k_\theta/\omega$ corresponding to whispering gallery modes with low disc-integrated visibilities.

In the following, the properties of each of the three types of phase-space structures are described in more detail.
Considering them separately is relevant for the study of gravito-inertial modes
because, according to Percival's conjecture \citep{P73,BR84} and following the results obtained for acoustic modes \citep{LG2},
we expect that each type of phase-space structure can be asymptotically related to families of modes with specific properties.

\subsubsection{Chaotic regions}
\label{cha}

PSS computed at different rotation rates show that a large chaotic region is systematically present
in the upper part of the $\left[0,N^{\rm p}_{0,{\rm max}}\right]$ frequency interval.
In the following we argue that this chaotic region is related to
the existence of unstable periodic orbits, which are themselves a direct consequence
of the presence of a local minimum of the Brunt-V\"{a}is\"{a}l\"{a} frequency in the stellar envelope.
This interpretation is supported by our ability to predict the frequency range for which
large chaotic regions are seen on the PSS.

In smoothly perturbed integrable Hamiltonian systems, chaotic motions appear near separatrices that connect hyperbolic points, that is unstable periodic orbits or unstable equilibrium points. In the framework of the KAM transition theory, these hyperbolic points naturally originate from 
the destruction of the resonant tori and chaos first appears around these hyperbolic points and along the structures that connect them \citep{O02}.
Chaotic motions, however, can also originate from hyperbolic points already present in the unperturbed system.
In our case, the unperturbed system does  possess hyperbolic points as can be inferred from the PSS at $\Omega=0$ shown in panel (a) of Fig.~\ref{SdPsep}.
The vicinity of the point $(k_r, r) = (0,0.8)$ is indeed typical of the phase portrait around a hyperbolic point with nearby orbits
forming branches of hyperbolas.
The separatrix (the red curve in panel (a) of Fig.~\ref{SdPsep}) is also called a homoclinic orbit because it connects the hyperbolic point to itself.
The apex of the motion of a pendulum provides a classical example of a hyperbolic point in a system with one degree of freedom, which is in this case a hyperbolic fixed point or equivalently
an unstable equilibrium point.
In our system with two degrees of freedom, the hyperbolic point is rather an unstable periodic orbit because it corresponds to a circular trajectory at a 
fixed radius ($r\sim0.8$ on panel (a) of Fig.~\ref{SdPsep}).
Since the ray dynamics in a spherical star is separable in the radial and latitudinal coordinates, the radius of the hyperbolic point is
actually a fixed point of the radial ray dynamics.
As in the case of the pendulum, the radial Hamiltonian takes the usual form $H_r = 1/2 k_r^2 + V_r$ and the 
unstable fixed point corresponds to the maximum of the potential $V_r$.

In Appendix~\ref{sec:hyperbolic}, analytical solutions of this problem are found via
the envelope approximation, which involves the assumption\ that the mass enclosed in a sphere of given $r$ does not depend on $r$, and
 we show that these solutions provide a very good approximation to the numerical solutions obtained without the envelope approximation.
Accordingly, the gravity ray dynamics possesses a hyperbolic point for any frequency $\omega$ within the
$[0.93 N_{0,{\rm min}}, N_{0,{\rm min}}]$ interval. The radius of the hyperbolic point varies monotonically from $r=0.84 R$ for $\omega=0.93 N_{0,{\rm min}}$
to $r=0.75 R$ for $\omega=N_{0,{\rm min}}$, while the value of $L$ characterising the separatrix varies also monotonically from $L=16.92$ for
$\omega=0.93 N_{0,{\rm min}}$
to $L=+\infty$ for $\omega=N_{0,{\rm min}}$.

\begin{figure}
\resizebox{\hsize}{!}{\includegraphics{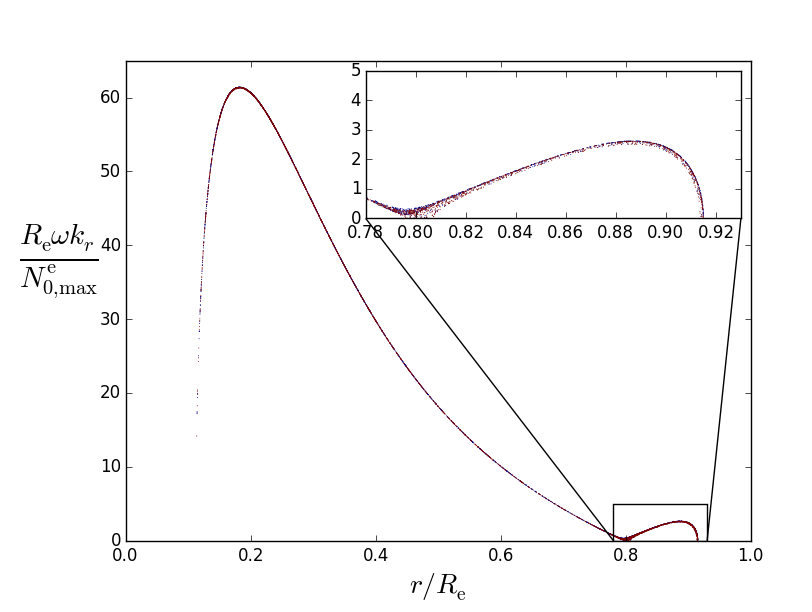}}
\caption{PSS computed at  $\omega/f=30.5$ for a rotation $\Omega/\Omega_{\rm K} = 0.02$ showing the development of a small chaotic zone
in the vicinity of the separatrix present in the non-rotating case (red curve on panel (a) of Fig.~\ref{SdPsep}).}
\label{chaos}
\end{figure}

To confirm that chaos first occurs around the separatrix,
we computed gravito-inertial rays close to this orbit for a slowly rotating model $\Omega/\Omega_{\rm K}=0.02$.
As illustrated by Fig.~\ref{chaos}, a small chaotic region is found in the expected frequency range and at the expected location.
 We further make the hypothesis that
the large chaotic regions observed at $\Omega/\Omega_{\rm K}=0.38$ and $\Omega/\Omega_{\rm K}=0.84$ are due to the same mechanism.
To test this hypothesis, PSS have been computed
for frequencies inside and outside
$\left[0.93 \sqrt{(N^{\rm e}_{0,{\rm min}})^2+ f^2}, N^{\rm p}_{0,{\rm min}}\right]$.
To define the lower and upper bounds of this interval,
rotational effects 
have been taken into account on phenomenological grounds.
First, rotation changes the frequency range for propagation, and following the conditions~\eq{approx1}, a $f^2 \sin^2 \Theta$ term must be
added to $N^2_{0,{\rm min}}$.
Second, we included the $\left[N^{\rm e}_{0,{\rm min}}, N^{\rm p}_{0,{\rm min}}\right]$ interval opened by the centrifugal force.
Indeed, we expect that unstable periodic orbits exist for these frequencies because there is always a range of latitudes
where the condition $\omega/\sqrt{N_{0,{\rm min}}^2+f^2\sin^2\Theta} \in [0.93, 1]$
is fulfilled. 
Considering frequencies 5\% apart from the interval limits,
we found that the large chaotic region is present inside and absent outside the $\left[0.93 \sqrt{(N^{\rm e}_{0,{\rm min}})^2+ f^2}, N^{\rm p}_{0,{\rm min}}\right]$ interval.

At a given frequency, the extent of the chaotic surface seen on the PSS is not simple to predict quantitatively. 
Comparing Figs.~\ref{chaos}, \ref{SdP31}, and \ref{SdP20} we nevertheless observe that, as expected, 
the surface of the chaotic region increases with rotation.

\subsubsection{Island chains}
\label{sec:island}

As they originate from the destruction of given rational tori, island chains 
can be associated with the rational number $p/q$ of their parent torus.
Then, the central periodic orbit of the $p/q$ island imprints $q$ times the PSS \citep{O02}; in this study we call the number $q$  the period.
The number $p$ can be found by constructing a different PSS (for instance by fixing the radius $r$ instead of the colatitude).
For super-inertial rays, $p$ corresponds to the number of loops visible in real space.
On the six PSS displayed in Figs.~\ref{SdP31}--\ref{SdP05}, we highlighted island chains of various periods:
one period-1 at $\omega/f =2.4$, 
two of period-5 (but with different $p$ values) 
at $\omega/f =3.1$ and $\omega/f =1.5$, respectively, one period-9 at $\omega/f =2$, one period-11 at $\omega/f =0.8$, and one with a period larger than 26 at $\omega/f=0.5$.
As expected from theory, the width of the island decreases with their period.
Moreover, there is a clear tendency to find only small and large-period island chains in the sub-inertial regime,
although there is no obvious change of behaviour at $\omega/f=1$.

In a typical KAM transition, island chains form immediately for a non-zero pertubation, grow with the amplitude of the perturbation, and are finally destroyed.
We followed the evolution of some low-period island chains with rotation and found that 
their width slightly increase but no destruction of island chains above a certain rotation rate has been observed.
For example, the period-1 island chain is still clearly visible at 99\% of the critical velocity.

In the physical space, rays belonging to island chains do not fill their resonant cavity as they remain concentrated around
the stable periodic trajectory.
From \citet{BL12}, we already know that numerically computed modes can be associated with island chains.
A detailed confrontation of the island properties with numerically computed modes will be performed in a future paper.

\subsubsection{Surviving tori}

The irrational tori of the non-rotating case are continuously deformed by the effect of rotation but may nevertheless have survived destruction
even at high rotation rates. This is  the case, in particular, for high-$L$ tori, since for most frequencies no large island chain or chaotic region is observed
in the PSS in this domain,\ which corresponds to high values of the normalised $k_r$. 
Some irrational tori with lower L also survive the perturbation induced by rotation.
This is obvious at low frequencies where phase space is very much structured by one-dimensional curves
broadly similar to the non-rotating curves (see Figs.~\ref{SdP08} and \ref{SdP05} and panel (c) of Fig.~\ref{SdPsep}).  

This nearly integrable behaviour of the low-frequency, gravito-inertial, ray dynamics might be attributed to the proximity of a hidden integrable
system. To test whether the ray dynamics in the traditional approximation could be this integrable system,
we computed the "traditional" invariant $\lambda$ along the path of gravito-inertial rays. 
Considering a sub-inertial ray in a $\Omega/\Omega_{\rm K} = 0.38$ star, we found that $\lambda$ varies by several orders of magnitude along a given ray. 
As expected, the largest variations occur
in regions of space that are outside the "traditional" domain of propagation, but even within the resonant cavity of the traditional approximation, $\lambda$ still varies
by a factor of two. 
In the super-inertial regime, the variations of $\lambda$ are of the order of $1.5$, which is of the same order of magnitude 
as the variations of $L$; the parameter $L$ is the invariant of the non-rotating case, which is not supposed to provide a valid approximation.
We conclude that the ray dynamics computed in the traditional approximation does not provide an accurate approximation 
even in the low-frequency regime, where the full gravito-inertial ray dynamics is close to integrable.

\section{From rays to modes}
\label{sec:modes}

In this section, we summarise the properties of gravito-inertial modes that can be inferred from the present study of the ray dynamics.
Methods to construct modes from the superposition of positively interfering rays have been first
proposed for integrable systems. The EBK procedure has been established in the field of quantum physics as a generalisation of the Bohr-Sommerfeld method 
to quantise systems from their classical trajectories. This method only works in the presence of invariant tori and leads to 
a regular frequency spectrum in the sense that the spectrum can be described by a function of N integers, where N is the number of degrees of freedom of the system. For chaotic dynamical systems, the EBK quantisation
does not work and must be replaced by the
Gutzwiller trace formula, which is difficult to apply in practice \citep{G90}. 
Nevertheless, generic statistical properties of spectra associated with chaotic dynamical systems, 
such as the distribution of spacings between consecutive frequencies, 
have been discovered in quantum systems \citep{BG84}. 
For mixed dynamical systems, it has been conjectured that chaotic regions and phase-space regions
structured by invariant tori, such as island chains or regions of surviving irrational tori, can be quantised separately.
Consequently, the spectrum of mixed systems is expected to be a superposition of frequency subsets, each of which are
associated with different phase-space structures. Frequency subsets associated with regions
structured by invariant tori should be regular while those associated with chaotic regions should possess generic
statistical properties \citep{P73,BR84}. A more detailed description of the concepts presented above together with the full references is available in \citet{LG2}.
Percival's predictions have been applied to acoustic rays in rotating stars and successfully confronted with
numerically computed high-frequency acoustic modes \citep{LG2}. 

In our case, Percival's conjecture implies that the frequency spectrum of short-wavelength gravito-inertial modes is a superposition of
frequency subsets associated with the phase-space structures of the gravito-inertial ray dynamics, notably a regular sub-spectrum associated
with the surviving irrational tori, regular sub-spectra associated with island chains, and a sub-spectrum with generic
statistical properties associated with the large chaotic region.
In principle, quantitative predictions on the regular sub-spectra can be obtained by conducting the EBK quantisation of the invariant tori.
This has been carried out in \citet{PL12} for the period-2 island chain of the acoustic ray dynamics and could be also attempted 
for the gravito-inertial island chains and surviving tori.

Previous numerical explorations of gravito-inertial modes already provide some insights into the relation between the asymptotic predictions of the gravito-inertial
dynamics and the numerically computed modes.
Using $\mu=3$ polytropic models, \citet{BL10} followed low-degree ($\ell \le 3$) modes in both low- and high-order ranges 
from $\Omega=0$ to $\Omega=0.7 \Omega_{\rm K}$. The vast majority of these modes undergo a smooth evolution of their spatial distribution with rotation, in which
the main change is the equatorial mode confinement in the sub-inertial regime. These modes most probably correspond to the phase-space region
structured by the surviving irrational tori. On the other hand, some modes experience a striking modification of their spatial distribution,
leading to a concentration of their energy. The connection between these modes called rosette modes and the central periodic orbits of island chains has been clearly established in \citet{BL12}.
Since then, \citet{TS13} proposed a different approach in which rosette modes result from the rotationally induced coupling of modes, which were
nearly degenerate gravity modes at $\Omega=0$ \citep[see also][]{ST14,TT14}.
The link between these two apparently very different approaches to model rosette modes remains to be understood. 
In our study, the number $K$ introduced by \citet{TS13} to characterise families of rosette modes corresponds to the ratio $2q/p$.\footnote{When comparing modes and rays, one must be careful about the fact that all modes are either symmetric or anti-symmetric with respect to the equator, which is not the case for ray trajectories.}
The authors found only integer values of $K$, which they justified by the fact that the only effect of rotation they considered was the Coriolis force.
We also present  island chains with non-integer values.
This is probably because we took the centrifugal deformation into account.
Modes that could be related to the large chaotic region that we describe have not been identified yet.
By providing an estimate of both the frequencies and wave vectors of chaotic rays,
ray dynamics will help to find these modes.
More generally, a dedicated numerical study will be necessary to test the predictions of the ray-based asymptotic theory.
In this context, tools allowing us to construct phase-space representations for numerically computed modes, such as Husimi distributions, are available.
As illustrated by \citet{LG2}, those can be used to show the imprint of modes on the PSS.

\section{Discussion and conclusions}
\label{sec:discussion}

The main results presented in the previous sections are (i) the derivation
of an eikonal equation for gravito-inertial rays in polytropic uniformly rotating stars, (ii) the analysis of the properties of this equation,
including constraints on the domains of propagation for $k_\phi=0$  rays, (iii) the numerical computation of the $k_\phi=0$ ray dynamics
for $\mu=3$ polytropic models with increasing rotation, and (iv) the interpretation of the numerical results using tools and concepts of Hamiltonian dynamics.
In particular, the exploration of the phase space provides us with a global view of the properties of gravito-inertial waves in rotating stars.
Qualitative predictions on the frequency spectrum of 
gravito-inertial modes have been made and should be a powerful tool to interpret the properties of numerically computed gravito-inertial modes.
A quantitative result already provided by the ray theory is the domain of propagation. Other quantitative predictions
are expected from the EBK quantisation of the surviving irrational tori and of the island chains.

A first restriction of the present work is that we did not consider $k_\phi \ne 0$ rays. This would be necessary 
for an asymptotic study of non-axisymmetric gravito-inertial modes.
Another restriction is the use of $\mu=3$ polytropic models although the formalism can be extended to more realistic rotating stellar models.
In particular, the effects of a convective core and/or a convective envelope would be interesting to consider.
 
The present ray dynamics formalism can be improved 
by taking the effect of background mean motions and, in particular, differential rotation  into account.
For example, self-consistent models including baroclinic effects computed by the ESTER code \citep{REL13} could be used.
According to atmospheric or oceanic studies, the eikonal equation including
mean horizontal motions is identical
to (\ref{disp0}),
except that
the frequency is changed into the Doppler-shifted frequency, which is the frequency that would be observed in
a frame of reference moving with the mean horizontal motions \citep{FA03}. Analagous with the effect of wind shear, differential rotation should cause
some back-refraction and thus will modify the resonant cavity.
Another improvement to the eikonal equation would be to include non-axisymmetric Rossby waves.

Ray models might also be used to study the transport of angular momentum by waves.
This would require us to extend the ray 
formalism to treat the evolution of the wave amplitude $A$ along the ray path (the next order of the WKB approximation). 
Such amplitude equations are already used to investigate the driving of the atmospheric circulation by waves \citep{ME95}.
To conduct similar investigations in stars, 
it will be 
necessary to model energy deposition processes such as thermal dissipation and critical layers \citep{FV98, AM13}
and to take phase changes at caustics into account \citep{BR04}.

Angular momentum transport by gravito-inertial waves that occasionally escape the star
has been invoked as a possible explanation of the Be phenomenon \citep{IS13}. Specific properties of chaotic gravito-inertial waves 
such as those observed in our calculations could be interesting in this context. 
At any location and in particular at the surface, the horizontal wavelength $k_\perp$ of chaotic waves can indeed take many different values.
Thus a chaotic wave can be back-refracted many times at the surface before it escapes the star because $k_\perp$ has reached a high enough value.

When compared to the ray theory of \citet{DR00}, the present work includes the effects of centrifugal deformation as well as a realistic back-refraction of rays in the stellar envelope.
This last point has a crucial effect on the comparison between the two approaches. In \citet{DR00},
the rays of characteristic are confined into the star by a rigid and stress-free surface and this boundary condition
destroys the Hamiltonian
character of the dynamics of characteristics.
As a consequence, the volume of phase-space elements is not conserved, 
which in turns enables rays to focus on
attractors. This focusing effect is at the origin
of the singular gravito-inertial
modes that are effectively observed in the presence of rigid boundaries 
\citep{MB97,DR99}. However, there are no such rigid boundaries in real stars and singular gravito-inertial modes focused on attractors of characteristics by unrealistic rigid boundaries should not exist in stars.
As we have seen in the present paper, the ray dynamics ought to be Hamiltonian in stars as long as dissipative processes are neglected.

\begin{acknowledgements}
    The authors thank Michel Rieutord for his careful reading of the manuscript, and an anonymous referee for contributing to the improvement of this paper.
This work was supported by the Centre National
de la Recherche Scientifique (CNRS) through the Programme National de
Physique Stellaire (PNPS).
\end{acknowledgements}

\bibliographystyle{aa}
\bibliography{grav}

\appendix
\onecolumn
\newpage

\section{Derivation of the perturbation equation \eq{great0}}
\label{sec:deriv}

In this appendix, our goal is to transform the perturbation equations \eq{div}--\eq{adia} into
Eq.~\eq{great0} for $\hat \Psi$. In Sect.~\ref{sec:pressure}, density and velocity perturbations appearing in Eqs.~
\eq{div}--\eq{adia} are eliminated to obtain an equation for pressure perturbations only. Then in Sect.~\ref{sec:normal}, the new
variable $ \hat \Psi =  \hat P/a$ is introduced to eliminate the first-order derivatives in the meridional plane from the perturbation equation on $\hat P$.
Finally, in Sects.~\ref{sec:const} and \ref{sec:const2}, the
constant term ${\cal C}$ is expanded in powers of $c_{\rm s}^2$ to determine the dominant $\mathcal{O}(1/H^2_{\rm s})$ approximation of ${\cal C}$.

\subsection{Derivation of an equation for pressure perturbations only}
\label{sec:pressure}

We first rewrite the perturbation equations \eq{div}--\eq{adia}, using $\vec{u}=\rho_0 \vec{v}$ and the definition of the Brunt-V\"{a}is\"{a}l\"{a} frequency
$N_0^2= \vec{g}_0 \cdot \lp \frac{\nab \rho_{0}}{\rho_{0}} - \frac{1}{\Gamma_1}\frac{\nab P_0}{P_0} \rp$ as follows:
\label{ann_eq_fin}
\begin{align}
{\partial}_t \rho + \nab \cdot \vec{u}      &=0,\label{continuite} \\
{\partial}_t \vec{u} + \vec f\wedge\vec u   &=
- \nab P + \rho \vec{g}_0, \label{qte_mvmt_pert}    \\
{\partial}_t P                              &= c_{\rm s}^{2} \lp {\partial}_t \rho + \frac{N_0^2}{g_0^2} \vec{g}_0 \cdot \vec{u} \rp. \label{isentr_N_0}
\end{align}
In order to eliminate $\vec{u}$ from these equations, we first calculate
$\vec f\wedge \eq{qte_mvmt_pert}$ as
\begin{equation}
\vec f\wedge\partial_t\vec u+(\vec f\cdot\vec u)\vec f-f^2\vec u=-\vec f\wedge\grad P+\rho\vec f\wedge\vec g_0.
\end{equation}
The first term on the LHS is then developed using $\partial_t \eq{qte_mvmt_pert}$ to give
\begin{equation}
\label{dttu}
\partial_{tt}^2\vec u+f^2\vec u-(\vec f\cdot\vec u)\vec f=-\grad\partial_tP+\partial_t\rho\vec g_0+\vec f\wedge\grad P-\rho\vec f\wedge\vec g_0.
\end{equation}
We use a similar method to eliminate the scalar product $\vec f\cdot\vec u$. We
compute~$\vec f\cdot \eq{qte_mvmt_pert}$
\begin{equation}
\vec f\cdot\partial_t\vec u=-\vec f\cdot\grad P+\rho\vec f\cdot\vec g_0,
\end{equation}
and introduce it into $\partial_t \eq{dttu}$ to get
\begin{equation}
\label{L_u}
\partial_{ttt}^3\vec u+f^2\partial_t\vec u=-(\vec f\cdot\grad P)\vec f+\rho(\vec f\cdot\vec g_0)\vec f -\grad\partial_{tt}^2P+\partial_{tt}^2\rho\vec g_0 + \vec f\wedge\grad\partial_t P-\partial_t\rho\vec f\wedge\vec g_0.
\end{equation}
Defining the operator
\begin{equation}
\label{def_L}
\mathcal{L} \equiv \partial_{ttt}^3+f^2\partial_t
\end{equation}
enables us to rewrite Eqs.~\eq{continuite} and \eq{isentr_N_0} as
\begin{align}
\partial_t\mathcal{L}(\rho)     &=      -\grad\cdot\mathcal{L}(\vec u),     \label{L_continuite}    \\
\partial_t\mathcal{L}(P)        &=      {c_{\rm s}}^2\partial_t\mathcal{L}(\rho)+\alpha\vec g_0\cdot\mathcal{L}(\vec u),   \label{L_isentr}
\end{align}
where $\mathcal{L}(\vec u)$ is given by Eq.~\eq{L_u} and $\alpha$, defined by
\begin{equation}
    \alpha=\frac{{c_{\rm s}}^2{N_0}^2}{{g_0}^2}=\frac{\mu\Gamma_1}{\mu+1}-1,
\end{equation}
is uniform in a polytropic model.
Replacing $\mathcal{L}(\vec u)$ in Eq.~\eq{L_isentr} by the RHS of Eq.~\eq{L_u}, we get
\begin{equation}
\label{rho_P}
\partial_t\mathcal{L}(P)+\alpha\left[(\vec f\cdot\vec g_0)(\vec f\cdot\grad P)+(\vec f\wedge \vec g_0)\cdot\grad\partial_t P+\vec g_0\cdot\grad\partial_{tt}^2P\right]
=      {c_{\rm s}}^2\partial_t\mathcal{L}(\rho)+\alpha\left[(\vec f\cdot\vec g_0)^2\rho+{g_0}^2\partial_{tt}^2\rho\right]={c_{\rm s}}^2\mathcal{D}(\rho),
\end{equation}
where
\begin{equation}
\mathcal{D}\equiv\partial_{tttt}^4+(f^2+{N_0}^2)\partial_{tt}^2+\frac{{N_0}^2(\vec f\cdot\vec g_0)^2}{{g_0}^2}.
\end{equation}

To simplify calculations, we assume that $P$ and $\rho$ are harmonic in time, i.e. $P=\Re\lbrace\hat Pe^{-i\omega t}\rbrace$ and $\rho=\Re\lbrace\hat\rho e^{-i\omega t}\rbrace$.
Then the operator $\mathcal{D}$ is simply a multiplication by
\begin{equation}
\label{expr_D}
D=\omega^4-\omega^2(f^2+{N_0}^2)+\frac{{N_0}^2(\vec f\cdot\vec g_0)^2}{{g_0}^2},
\end{equation}
and $\partial_t\mathcal{L}$ a multiplication by
\begin{equation}
G=\omega^2(\omega^2-f^2).
\end{equation}

From Eq.~\eqref{rho_P}, $\hat\rho$ is expressed as a function of $\hat P$ as follows:
\begin{equation}
\label{hat_rho}
\hat\rho=\frac{G\hat P+\vec H\cdot\grad\hat P}{{c_{\rm s}}^2D},
\end{equation}
with
\begin{equation}
\vec H=\alpha\left[(\vec f\cdot\vec g_0)\vec f-i\omega\vec f\wedge\vec g_0-\omega^2\vec g_0\right].
\end{equation}
We then compute $\grad\cdot\mathcal{L}(\vec u)$ from Eq.~(\ref{L_u}), using the general identity $\grad\cdot(b\vec a)=\vec a\cdot\grad b+b\grad\cdot\vec a$ and the fact that $\vec f$ is uniform
\begin{equation}
\label{div_L_u}
\grad\cdot\mathcal{L}(\vec u)=-\vec f\cdot\grad(\vec f\cdot\grad\hat P)-i\omega\grad\cdot(\vec f\wedge\grad\hat P)+\omega^2\Delta\hat P+\grad\cdot(\hat\rho\vec F),
\end{equation}
with
\begin{equation}
\vec F=(\vec f\cdot\vec g_0)\vec f+i\omega\vec f\wedge\vec g_0-\omega^2\vec g_0.
\end{equation}
The expression~(\ref{div_L_u}) can be simplified using
\begin{equation}
\grad\cdot(\vec f\wedge\grad\hat P)=\grad\hat P\cdot(\grad\wedge\vec f)-\vec f\cdot(\grad\wedge\grad\hat P)=0,
\end{equation}
which finally yields
\begin{equation}
\label{div_L_u_fin}
\grad\cdot\mathcal{L}(\vec u)=\omega^2\Delta\hat P-\vec f\cdot\grad(\vec f\cdot\grad\hat P)+\grad\cdot(\hat\rho\vec F).
\end{equation}
Replacing the expression~(\ref{div_L_u_fin}) in Eq.~(\ref{L_continuite}) and using Eq.~(\ref{hat_rho}) to express $\hat\rho$, we obtain the following equation
for $\hat P$:
\begin{equation}
\frac{G(G\hat P+\vec H\cdot\grad\hat P)}{{c_{\rm s}}^2D}+\omega^2\Delta\hat P-\vec f\cdot\grad(\vec f\cdot\grad\hat P)+\grad\cdot\left[\frac{(G\hat P+\vec H\cdot\grad\hat P)\vec F}{{c_{\rm s}}^2D}\right]=0.
\label{firstP}
\end{equation}

We use the identity $\grad\cdot(\vec a\wedge\vec b)=\vec b\cdot(\grad\wedge\vec a)-\vec a\cdot(\grad\wedge\vec b)$, which implies that $\grad\cdot(\vec f\wedge\vec g_0)=0$, as $\vec f$ is uniform and $\vec g_0$ derives from a potential. Defining the vector $\vec F_0 = (\vec f\cdot\vec g_0)\vec{f} -\omega^2\vec g_0$,
the vectors $\vec{F}$ and $\vec{H}$ are written as $\vec F = \vec F_0 + i \omega \vec f\wedge\vec g_0$ and $\vec H=\alpha (\vec F_0 - i \omega \vec f\wedge\vec g_0)$. Thus, Eq.~\eq{firstP} becomes
\begin{equation}
\begin{aligned}
   \frac{G(G\hat P+\alpha \vec F_0\cdot\grad\hat P)}{{c_{\rm s}}^2D}+\omega^2\Delta\hat P-\vec f\cdot\grad(\vec f\cdot\grad\hat P)+&\grad\cdot\left[\frac{(G\hat P+\alpha \vec F_0\cdot\grad\hat P)\vec F_0}{{c_{\rm s}}^2D}\right]
+ i(1-\alpha) \frac{\omega G}{c_{\rm s}^2 D} (\vec f\wedge\vec g_0)\cdot\grad\hat P \\
&\quad - i\alpha \omega \grad\cdot\left[(\vec f\wedge\vec g_0)\cdot\grad\hat P \frac{\vec F}{c_{\rm s}^2 D}\right] + i\omega\frac{\alpha}{c_{\rm s}^2D}(\vec f\wedge\vec g_0)\cdot\grad(\vec F_0\cdot\grad\hat P) =0,
\end{aligned}
\end{equation}
where we have used the fact that $\alpha$ is uniform and $\vec f\wedge\vec g_0$ is parallel to $\vec e_\phi$.
We then use the relation
\begin{equation}
\vec F_0\cdot\grad\left(\frac{\vec F_0}{{c_{\rm s}}^2D}\cdot\grad\hat P\right)
 =      \frac{1}{{c_{\rm s}}^2D}\vec F_0\cdot\grad(\vec F_0\cdot\grad\hat P)+\left[\vec F_0\cdot\grad\left(\frac{1}{{c_{\rm s}}^2D}\right)\right]\vec F_0\cdot\grad\hat P
\end{equation}
to get
\begin{equation}
\label{eq_inter}
\begin{aligned}
D\omega^2\Delta\hat P-D\vec f\cdot\grad(\vec f\cdot\grad\hat P)+\frac{{N_0}^2}{{g_0}^2}\vec F_0\cdot\grad(\vec F_0\cdot\grad\hat P)+\frac{N}{{c_{\rm s}}^2}\vec F_0\cdot\grad\hat P+\frac{M}{{c_{\rm s}}^2}\hat P
+ &i(1-\alpha) \frac{\omega G}{c_{\rm s}^2} (\vec f\wedge\vec g_0)\cdot\grad\hat P - i\alpha \omega D \grad\cdot\left[(\vec f\wedge\vec g_0)\cdot\grad\hat P \frac{\vec F}{c_{\rm s}^2 D}\right]\\
&\quad +i\omega\frac{\alpha}{c_{\rm s}^2}(\vec f\wedge\vec g_0)\cdot\grad(\vec F_0\cdot\grad\hat P) = 0,
\end{aligned}
\end{equation}
where
\beqa
M       &=&      G\left[G+{c_{\rm s}}^2D\grad\cdot\left(\frac{\vec F_0}{{c_{\rm s}}^2D}\right)\right],      \\
N       &=&      (1+\alpha)G+{c_{\rm s}}^2D\grad\left(\frac{\alpha\vec F_0}{{c_{\rm s}}^2D}\right).
\eeqa
From the definition of $\vec F_0$, we have
\begin{equation}
\begin{aligned}
\vec F_0\cdot\grad(\vec F_0\cdot\grad\hat P)        &=      (\vec f\cdot\vec g_0)^2\vec f\cdot\grad(\vec f\cdot\grad\hat P)+\omega^4\vec g_0\cdot\grad(\vec g_0\cdot\grad\hat P)
                                            -\omega^2(\vec f\cdot\vec g_0)\left[\vec f\cdot\grad(\vec g_0\cdot\grad\hat P)+\vec g_0\cdot\grad(\vec f\cdot\grad\hat P)\right]   \\
                                            &\quad  +\left[(\vec f\cdot\vec g_0)\vec f\cdot\grad(\vec f\cdot\vec g_0)-\omega^2\vec g_0\cdot\grad(\vec f\cdot\vec g_0)\right]\vec f\cdot\grad\hat P,
\end{aligned}
\end{equation}
where, thanks to the fact that $\vec f$ is uniform, the last term can be also written $\frac{1}{2}(\vec f\cdot\grad K)\vec f\cdot\grad\hat P$ with
\begin{equation}
K=(\vec f\cdot\vec g_0)^2-\omega^2{g_0}^2=\vec F_0\cdot\vec g_0.
\end{equation}
In the end, we get
\begin{equation}
\begin{aligned}
\label{eq_finale}
       D\Delta\hat P-(\omega^2-f^2-{N_0}^2)\vec f\cdot\grad(\vec f\cdot\grad\hat P)+\frac{{N_0}^2\omega^2}{{g_0}^2}\vec g_0\cdot\grad(\vec g_0\cdot\grad\hat P)
     -&\frac{{N_0}^2}{{g_0}^2}(\vec f\cdot\vec g_0)[\vec f\cdot\grad(\vec g_0\cdot\grad\hat P)+\vec g_0\cdot\grad(\vec f\cdot\grad\hat P)]        \\
\quad  +\frac{N\vec f\cdot\vec g_0+\frac{\alpha}{2}\vec f\cdot\grad K}{{c_{\rm s}}^2\omega^2}\vec f\cdot\grad\hat P-\frac{N}{{c_{\rm s}}^2}\vec g_0\cdot\grad\hat P+\frac{M}{{c_{\rm s}}^2\omega^2}\hat P
     + &i(1-\alpha) \frac{G}{\omega c_{\rm s}^2} (\vec f\wedge\vec g_0)\cdot\grad\hat P - i\alpha \frac{D}{\omega} \grad\cdot\left[(\vec f\wedge\vec g_0)\cdot\grad\hat P \frac{\vec F}{c_{\rm s}^2 D}\right]\\
     & \quad+i\frac{\alpha}{\omega c_{\rm s}^2}(\vec f\wedge\vec g_0)\cdot\grad(\vec F_0\cdot\grad\hat P) = 0,
\end{aligned}
\end{equation}
which is an equation for a single unknown, as we were looking for.

Using the relation $D\Delta \hat{P} = \omega^2\left[(\omega^2-f^2)\Delta \hat{P} - N_0^2 \Delta \hat{P} + \frac{N_0^2 (\vec f\cdot\vec g_0)^2}{\omega^2g_0^2}\Delta \hat{P}\right]$,
from the definition of $D$, Eq.~\eq{eq_finale}
is rewritten as
\begin{equation}
\begin{aligned}
\label{eq_finale_bis}
 (\omega^2-f^2)\Delta\hat P =
 &\frac{(\omega^2-f^2)}{\omega^2} \vec f\cdot\grad(\vec f\cdot\grad\hat P)+
 \frac{{N_0}^2}{\omega^2}\left\lbrace\vphantom{\frac{(\vec f\cdot\vec g_0)}{g_0^2}}\omega^2 \Delta \hat P - \frac{\omega^2}{{g_0}^2}\vec g_0\cdot\grad(\vec g_0\cdot\grad\hat P) - \vec f\cdot\grad(\vec f\cdot\grad \hat P ) \right.    \\
&\quad \left.-\frac{(\vec f\cdot\vec g_0)^2}{g_0^2} \Delta \hat P\frac{(\vec f\cdot\vec g_0)}{g_0^2} \left[\vec f\cdot\grad(\vec g_0\cdot\grad \hat P )+\vec g_0\cdot\grad(\vec f\cdot\grad \hat P )\right] \right\rbrace  -\frac{\vec{\cal V}}{\omega^2}\cdot\grad \hat P - \frac{M}{{c_{\rm s}}^2\omega^4} \hat P\\
&\quad -i(1-\alpha) \frac{G}{\omega^3 c_{\rm s}^2} (\vec f\wedge\vec g_0)\cdot\grad \hat P + i\alpha \frac{D}{\omega^3} \grad\cdot\left[(\vec f\wedge\vec g_0)\cdot\grad\hat P \frac{\vec F}{c_{\rm s}^2 D}\right]-i\frac{\alpha}{\omega^3 c_{\rm s}^2}(\vec f\wedge\vec g_0)\cdot\grad(\vec F_0\cdot\grad\hat P),
\end{aligned}
\end{equation}
where
\beqa
\vec{\cal V} = \frac{N\vec f\cdot\vec g_0+\frac{\alpha}{2}\vec f\cdot\grad K}{{c_{\rm s}}^2\omega^2}\vec f - \frac{N}{{c_{\rm s}}^2}\vec g_0.
\eeqa
To greatly simplify the terms involving second-order derivatives, one can express $\grad\hat P$ in the (non-orthonormal) basis defined by $\vec f$, $\vec g_0$, and $\vec e_\phi$ and take the divergence of the resulting equation. It gives
\beqa
-g_0^2\n_{\!\!\vec{f}}^2 + \vec f\cdot\vec g_0 \lp \n_{\!\!\vec g_0 }\n_{\!\!\vec{f}} + \n_{\!\!\vec{f}} \n_{\!\!\vec g_0 }\rp - f^2 \n_{\!\!\vec g_0 }^2 = -\left[f^2g_0^2 - (\vec f\cdot\vec g_0)^2\right] \lc \Delta - \frac{1}{(r \sin\theta)^2}\frac{\partial^2}{\partial\phi^2 } \rc  + \lp T' \vec g_0 + R' \vec f \rp \cdot \grad,
\eeqa
where we define the following operators and expressions:
\beqa
\label{gr1}
\n_{\!\!\vec{f}}^2 & \equiv & \n_{\!\!\vec{f}} (\n_{\!\!\vec{f}} ) \qquad \;  \mbox{with} \;\;\;\; \n_{\!\!\vec{f}} \equiv \vec{f} \cdot \nab, \\
\n_{\!\!\vec g_0 }^2 & \equiv & \n_{\!\!\vec g_0 } (\n_{\!\!\vec g_0 } ) \qquad \;  \mbox{with} \;\;\;\; \n_{\!\!\vec g_0 } \equiv \vec g_0  \cdot \nab,\\
T'&=& -\n_{\!\!\vec{f}}p + \n_{\!\!\vec g_0 }n + \frac{p}{\sigma} \n_{\!\!\vec{f}}\sigma -\frac{n}{\sigma} \n_{\!\!\vec g_0 }\sigma - p \grad \cdot \vec f + n \grad \cdot \vec g_0, \\
R'&=& -\n_{\!\!\vec g_0 }p + \n_{\!\!\vec{f}}q - \frac{q}{\sigma} \n_{\!\!\vec{f}}\sigma +\frac{p}{\sigma} \n_{\!\!\vec g_0 }\sigma + q \grad \cdot \vec f - p \grad \cdot \vec g_0,
\eeqa
with $p = \vec f\cdot\vec g_0$, $q= g_0^2$, $n=f^2$ and $\sigma=(\vec f\cdot\vec g_0)^2 - f^2 g_0^2$.
If in addition, the operators $\n_z$ and $\Delta_{\perp}$ are defined as in Eqs.~(\ref{gr0}) and (\ref{gr00}), the wave equation (\ref{eq_finale_bis}) becomes
\beqa
\label{good}
\Delta\hat P = \frac{f^2}{\omega^2}\n_z^2\hat P + \frac{N_0^2}{\omega^2} \Delta_{\perp}\hat P + (\vec{\cal V}' + i {\vec{\cal V}'}_{\!\!\!\!m})\cdot\grad \hat P + M' \hat P,
\eeqa
where, using the fact that $\vec f\wedge\vec g_0=-fg_0\sin\Theta\vec e_\phi$,
\begin{align}
\vec{\cal V}' &= \frac{N_0^2}{\omega^2(\omega^2-f^2)g_0^2}\lp T'\vec g_0 + R' \vec f \rp  - \frac{\vec{\cal V}}{\omega^2(\omega^2-f^2)} + \frac{N_0^2}{\omega^2}\lp - \frac{\n_{\parallel} g_0 }{g_0} + \grad \cdot \vec e_{\parallel}  \rp \vec e_{\parallel} , \label{v0} \\
{\vec{\cal V}'}_{\!\!\!\!m} &=  - \frac{1}{c_{\rm s}^2 \omega} \lc (1-\alpha) - \alpha \frac{c_{\rm s}^2 D}{G} \grad \cdot \lp \frac{\vec F_0}{c_{\rm s}^2 D}\rp - \alpha \frac{r \sin\theta}{G g_0 \sin\Theta} \vec F_0 \cdot \nab \lp \frac{g_0 \sin\Theta}{r \sin\theta} \rp \rc \vec{f} \wedge \vec g_0,\label{vm} \\
M' &= - \frac{M}{{c_{\rm s}}^2\omega^4(\omega^2-f^2)}. \label{mm}
\end{align}

\subsection{Normal form}
\label{sec:normal}

Here, Eq.~\eq{mm} for pressure perturbations is written in a normal form by eliminating first-order derivatives in the meridional plane. This can be done by introducing a new variable $\hat \Psi$ defined
by $\hat P = a \hat \Psi$, where
the axisymmetric function $a$ is chosen to eliminate the first-order term $\vec{\cal V}' \cdot\grad \hat P$. To carry out this substitution, we use the following relations:
\beqa
\Delta \hat P &=& a \Delta \hat \Psi + 2 \grad a \cdot \grad \hat\Psi + \hat\Psi\Delta a, \\
\n_{\parallel}^2 \hat P & =& a \n_{\parallel}^2 \hat\Psi + 2  (\n_{\parallel} a) (\n_{\parallel} \hat \Psi)  + \hat\Psi\n_{\parallel}^2 a, \\
\n_{z}^2 \hat P & =& a \n_{z}^2 \hat\Psi + 2  (\n_{z} a) (\n_{z} \hat \Psi)  + \hat\Psi\n_{z}^2 a,
\eeqa
where the operators $\n_{\parallel}$ and $\n_{z}$ are defined by Eqs.~(\ref{gr0}) and (\ref{gr00}).
Equation~\eq{good} becomes
\beqa
\label{toogood}
a \Delta \hat \Psi = a \frac{f^2}{\omega^2}\n_{z}^2\hat \Psi + a \frac{N_0^2}{\omega^2} \Delta_{\perp}\hat \Psi + \vec{\cal V}''\cdot\grad \hat \Psi +
i a {\vec{\cal V}'}_{\!\!\!\!m}\cdot\grad \hat \Psi + M'' \hat \Psi,
\eeqa
where
\beqa
\vec{\cal V}'' &=& - 2 \grad a + 2 \frac{f^2}{\omega^2} (\n_z a) \vec e_z + 2\frac{N_0^2}{\omega^2} \grad a - 2\frac{N_0^2}{\omega^2} (\n_{\parallel}a) \vec e_{\parallel} + a  \vec{\cal V}',\\
M'' &=& -\Delta a + \frac{f^2}{\omega^2} \n_z^2 a+ \frac{N_0^2}{\omega^2} \Delta a - \frac{N_0^2}{\omega^2} \n_{\parallel}^2 a - \frac{N_0^2}{\omega^2} (\grad \cdot \vec e_{\parallel}) \n_{\parallel} a + \vec{\cal V}' \cdot \grad a + M' a.
\eeqa

The condition $\vec{\cal V}''=\vec0$ translates into a first-order linear differential equation for $a$. The projection of this equation on $\vec e_{\parallel}$ and $\vec e_z$ leads
to two equations for $\n_{\parallel}a/a$ and $\n_z a/a$. These equations can be solved as a linear system for these two unknowns provided that
the determinant of the system, which
is equal to $D$, does not vanish. We then use these expressions of $\n_{\parallel}a/a$ and $\n_z a/a$ to write
\beqa
\label{eq:a}
\frac{\grad a}{a} = {\vec{\cal W}},
\eeqa
where
\beqa
\label{wpara}
{\vec{\cal W}} &=&{\cal W}_{\!\!\parallel}  \vec e_{\parallel} + {\cal W}_{\!\!z} \vec e_z,  \\
{\cal W}_{\!\!\parallel} &=& \frac{\omega^2}{2D} \left[(\omega^2 - N_0^2 - f^2) {\cal V'}_{\!\!\parallel}        - N_0^2 \cos \Theta {\cal V'}_{\!\!z} \right], \\
{\cal W}_{\!\!z} &=& \frac{\omega^2}{2D} \lp f^2 \cos \Theta {\cal V'}_{\!\!\parallel} + \omega^2 {\cal V'}_{\!\!z} \rp,
\eeqa
where ${\cal V'}_{\!\!\parallel}$ and ${\cal V'}_{\!\!z}$ are defined by $\vec{\cal V}' = {\cal V'}_{\!\!\parallel} \vec e_{\parallel} + {\cal V'}_{\!\!z} \vec e_z$.
With this choice for the function $a$, the wave equation Eq.~\eq{toogood} simplifies into
\beqa
\label{great}
 \Delta \hat \Psi = \frac{f^2}{\omega^2}\n_{z}^2\hat \Psi + \frac{N_0^2}{\omega^2} \Delta_{\perp}\hat \Psi + i \frac{\|\vec{f}\wedge\vec{g}_0\|}{\omega c_{\rm s}^2} {\cal T} \vec e_{\phi} \cdot \nab \hat \Psi + {\cal C} \hat \Psi,
\eeqa
where the ${\vec{\cal V}'}_{\!\!\!\!m}$ term is now expressed as a function of the dimensionless quantity ${\cal T}$, defined by
\begin{equation}
\label{T}
\begin{aligned}
{\cal T} &= (1-\alpha) - \alpha \frac{c_{\rm s}^2 D}{G} \grad \cdot \lp \frac{\vec F_0}{c_{\rm s}^2 D}\rp - \alpha \frac{r \sin\theta}{G g_0 \sin\Theta} \vec F_0 \cdot \nab \lp \frac{g_0 \sin\Theta}{r \sin\theta} \rp.
\end{aligned}
\end{equation}
The term ${\cal C}$ equals $M''/a$ and can be expressed in terms of $\vec{\cal W}$ to be written as
\begin{equation}
\label{eq:const}
\begin{aligned}
{\cal C} =& - \grad \cdot \vec{\cal W} - \vec{\cal W}^2 + \frac{N_0^2}{\omega^2} \left[ \grad \cdot \vec{\cal W} - \n_{\parallel} (\vec{\cal W}\cdot\vec e_{\parallel}) \right] + \frac{N_0^2}{\omega^2} \left[ \vec{\cal W}^2 - (\vec{\cal W}\cdot\vec e_{\parallel})^2 \right] - \frac{N_0^2}{\omega^2} (\grad \cdot \vec e_{\parallel}) (\vec{\cal W}\cdot\vec e_{\parallel})   \\
        &\quad + \frac{f^2}{\omega^2} \left[ \n_{z}  (\vec{\cal W}\cdot\vec e_z)  +  (\vec{\cal W}\cdot\vec e_z)^2 \right] + \vec{\cal V'} \cdot \vec{\cal W} + M' .
\end{aligned}
\end{equation}
From the definitions of $M'$ and $M$, it is easy to see that ${\cal C}$ can be written as ${\cal C} =-\omega^2/c_{\rm s}^2 + \mathcal{C'}$ to recover Eq.~\eq{great0}.

\subsection{The constant term}
\label{sec:const}

To take into account the back-refraction of outgoing waves in the  short-wavelength approximation,
${\cal C}$ is expanded in powers of $H={\cal R} T/g_0$ or equivalently $c_{\rm s}^2$ and we only retain
the dominant term.
Among the quantities involved in ${\cal C}$, the inverse of the sound speed and the Brunt-V\"{a}is\"{a}l\"{a} frequency tend
to become very large towards the stellar surface, whereas $\vec f ,\vec g_0$, and $\Gamma_1$ remain finite.
To derive the dominant term of ${\cal C}$ approaching the stellar surface, we use the following expression for $\vec{\cal W}$:
\beqa
\label{w0_0}
\vec{\cal W} =\frac{\vec w_0 + \vec w_1 c_{\rm s}^2}{c_{\rm s}^2 U},
\eeqa
where
\beqa
U = c_{\rm s}^2 D = \alpha K + G c_{\rm s}^2,
\eeqa
and $\vec w_0 = w_0 \vec e_{\parallel}$ and $\vec w_1$ are both ${\cal O}(1)$ with respect to the $c_{\rm s}^2$ expansion.

The dominant terms in the expression \eq{eq:const} of ${\cal C}$ are
\beqa
-\grad \cdot \vec{\cal W} + \frac{f^2}{\omega^2} \n_{z}  (\vec{\cal W}\cdot\vec e_z)  &=& \frac{\alpha\beta w_0 K^2}{\omega^2 g_0}  \frac{1}{c_{\rm s}^4 U^2},  \\
- \vec{\cal W}^2 + \frac{f^2}{\omega^2} (\vec{\cal W}\cdot\vec e_z)^2 + \frac{N_0^2}{\omega^2} \left[ \vec{\cal W}^2 - (\vec{\cal W}\cdot\vec e_{\parallel})^2 \right] &=& \frac{w_0^2 K}{\omega^2 g^2_0} \frac{1}{c_{\rm s}^4 U^2},         \\
\frac{N_0^2}{\omega^2} \left[ \grad \cdot \vec{\cal W} - \n_{\parallel} (\vec{\cal W}\cdot\vec e_{\parallel}) \right] &=& \frac{\alpha^2 g_0^2 K w_0 \grad \cdot \vec e_{\parallel}}{\omega^2} \frac{1}{c_{\rm s}^4 U^2},         \\
- \frac{N_0^2}{\omega^2} (\grad \cdot \vec e_{\parallel}) (\vec{\cal W}\cdot\vec e_{\parallel}) &=& - \frac{\alpha^2 g_0^2 K w_0 \grad \cdot \vec e_{\parallel}}{\omega^2} \frac{1}{c_{\rm s}^4 U^2}, \\
\vec{\cal V}' \cdot \vec{\cal W}   &=& -\frac{2 w_0^2 K}{\omega^2 g_0^2} \frac{1}{c_{\rm s}^4 U^2},
\eeqa
whereas $M' = {\cal O} (1/c_{\rm s}^2)$ is negligible.
Summing these terms we obtain, in the $K \ne 0$ case,
\beqa
{\cal C} = \frac{1}{\alpha^2 K^2} \frac{1}{c_{\rm s}^4} w_0 \lp \frac{\alpha \beta K^2}{\omega^2 g_0} + \frac{w_0 K}{\omega^2 g^2_0}+ \frac{\alpha^2 g_0^2 K \grad \cdot \vec e_{\parallel}}{\omega^2} - \frac{\alpha^2 g_0^2 K \grad \cdot \vec e_{\parallel}}{\omega^2} -\frac{2 w_0 K}{\omega^2 g_0^2} \rp = \frac{1}{\omega^2 c_{\rm s}^4}  \frac{w_0}{\alpha g_0} \lp \beta - \frac{w_0}{\alpha K g_0} \rp + {\cal O} (1/c_{\rm s}^2).
\eeqa
As shown below in Sect.~\ref{sec:const2}, the calculation of $w_0$ leads to
\beqa
w_0 = - \frac{\alpha K g_0}{2}(1+\alpha-\beta),
\eeqa
thus to
\beqa
{\cal C} = - \frac{K}{4 \omega^2 c_{\rm s}^4}(1+\alpha-\beta)(1+\alpha+\beta) + {\cal O} (1/c_{\rm s}^2) = \left(1 - \frac{f^2}{\omega^2} \cos^2 \Theta\right) \frac{\Gamma_1^2}{4} \frac{\mu-1}{\mu+1} \frac{g_0^2}{c_{\rm s}^4} + {\cal O} (1/c_{\rm s}^2).
\eeqa

In the sub-inertial regime ($\omega<f$), the dominant term of ${\cal C}$ vanishes at the critical angles $\Theta_{\rm c}$ and $\pi -\Theta_{\rm c}$
such that $\cos \Theta_{\rm c} = \omega/f$. Coming back to the full expression of ${\cal C}$, Eq.~\eq{eq:const}, it can be shown that 
\beqa
{\cal C} \simeq \frac{C_0 + C_1 c_{\rm s}^2 + C_2 c_{\rm s}^4 + C_3 c_{\rm s}^6}{(\alpha K + G c_{\rm s}^2)^2 c_{\rm s}^4}
\eeqa
and that $C_1$ does not vanish when $K=0$. 
Thus, as one approaches the surface along the critical angles, 
the dominant term in the expression of ${\cal C}$ is given by
\beqa
{\cal C} = \frac{C_1(\Theta_{\rm c})}{G^2} \frac{1}{c_{\rm s}^6(\Theta_{\rm c})} + {\cal O}(1/c_{\rm s}^6).
\eeqa
In practice, we found that rays do not reach the surface layers near the critical angle. This can be understood as the dispersion relation
and the ray equations show that near the critical angle, $k_{\perp}$ vanishes, and this implies that
${\rm d} r(\theta)/{\rm d}\theta$ vanishes as well.

\subsection{The $w_0$ term}
\label{sec:const2}

According to Eqs.~\eq{w0_0} and \eq{wpara}, $w_0$ is the dominant term of $c_{\rm s}^2U{\cal W}_{\!\!\parallel}$.
From the definitions of ${\cal W}_{\!\!\parallel}$ and $\vec{\cal V}'$, we obtain
\beqa
w_0 = \frac{\alpha \omega^2 g_0^2}{2 G} \lc \alpha g_0 T' -\alpha f \cos \Theta R' - \frac{K}{g_0 \omega^2} N + \frac{\alpha f \cos \Theta}{2 \omega^2} \vec f \cdot \grad K + \alpha g_0 (\omega^2 -f^2) (\n_{\parallel} g_0 - g_0 \grad \cdot \vec e_{\parallel}) \rc.
\eeqa
This expression is then developed using
\beqa
T' &=& - \vec f \cdot \grad (\vec f \cdot \vec g_0) + \left[ (\vec f \cdot \vec g_0) \vec f -f^2 \vec g_0 \right] \cdot \frac{\grad \sigma}{\sigma} + f^2 \grad \cdot \vec g_0, \\
R' &=& - \vec g_0 \cdot \grad (\vec f \cdot \vec g_0) + \vec f \cdot \grad g_0^2 -\frac{g_0^2}{\sigma} \vec f \cdot \grad \sigma + \frac{\vec f \cdot \vec g_0}{\sigma} \vec g_0 \cdot \grad \sigma - (\vec f \cdot \vec g_0) \grad \cdot \vec g_0,  \\
g_0 T' - f \cos \Theta R' &=& -\frac{1}{g_0} \lc (\vec f \cdot \vec g_0) \vec g_0 + g_0^2 \vec f \rc \cdot \grad (\vec f \cdot \vec g_0) + \frac{\vec g_0}{g_0} \cdot \grad \sigma - f \cos \Theta \vec f \cdot \grad g_0^2 - \frac{\sigma}{g_0} \grad \cdot \vec g_0,  \\
N       &=&      (1+\alpha)G+{c_{\rm s}}^2D\grad\left(\frac{\alpha\vec F_0}{{c_{\rm s}}^2D}\right) = (1+\alpha)G+\alpha \grad\cdot\vec F_0 -\frac{\alpha \vec F_0 \cdot \grad U}{ U},  \\
\grad U &=&  \alpha \grad K + \beta G \vec g_0, \\
\grad U \cdot \vec F_0 &=& \alpha \grad K\cdot \vec F_0 + \beta G K, \\
\grad\cdot\vec F_0  &=& \vec f \cdot \grad (\vec f \cdot \vec g_0) - \omega^2 \grad \cdot \vec g_0.
\eeqa
Thus, as $c_{\rm s}^2$ vanishes,
\beqa
\lim_{c_{\rm s}^2 \to 0} N =   (1+\alpha-\beta)G - \alpha \omega^2 \grad \cdot \vec g_0 - \alpha \frac{\grad K\cdot \vec F_0}{K} + \alpha \vec f \cdot \grad (\vec f \cdot \vec g_0).
\eeqa
The expression of $w_0$ becomes
\begin{align}
\
\frac{2 G}{\alpha \omega^2 g_0^2}w_0 &=
\begin{aligned}[t]
&-\frac{K}{g_0 \omega^2}(1+\alpha-\beta) G + \frac{\alpha K}{g_0} \grad \cdot \vec g_0 + \frac{\alpha}{g_0 \omega^2} \lc \grad K\cdot \vec F_0 -K \vec f \cdot \grad (\vec f \cdot \vec g_0) \rc + \alpha (g_0 T' - f \cos \Theta R') \\
&\quad+ \alpha \frac{f \cos \Theta}{2 \omega^2} \vec f \cdot \grad K + \alpha g_0 (\omega^2 -f^2) (\n_{\parallel} g_0 - g_0 \grad \cdot \vec e_{\parallel})
\end{aligned}   \\
&= -\frac{K}{g_0 \omega^2}(1+\alpha-\beta) G + \frac{\alpha}{g_0}(K-\sigma)\grad \cdot \vec g_0 + \frac{\alpha}{g_0} {\cal A},
\end{align}
where
\begin{equation}
\begin{aligned}
{\cal A} =& \frac{\grad K\cdot \vec F_0}{\omega^2} - \frac{K}{\omega^2} \vec f \cdot \grad (\vec f \cdot \vec g_0) - \left[ (\vec f \cdot \vec g_0) \vec g_0 + g_0^2 \vec f \right] \cdot \grad (\vec f \cdot \vec g_0) + \vec g_0 \cdot \grad \sigma - g_0 f \cos \Theta \vec f \cdot \grad g_0^2 + \frac{g_0 f \cos \Theta}{2 \omega^2} \vec f \cdot \grad K   \\
&\quad+ g_0^2 (\omega^2 -f^2) (\n_{\parallel} g_0 - g_0 \grad \cdot \vec e_{\parallel}).
\end{aligned}
\end{equation}
Developing the last expression, we show that
\beqa
{\cal A} = - (\vec f \cdot \vec g_0)\vec g_0 \cdot \grad (\vec f \cdot \vec g_0) +\frac{1}{2} (\vec f \cdot \vec g_0) \vec f \cdot\grad g_0^2 - g_0^2 (\omega^2 -f^2) (\n_{\parallel} g_0 + g_0 \grad \cdot \vec e_{\parallel}).
\eeqa
Thus,
\begin{align}
-\frac{2}{\alpha K g_0}w_0 &=
\begin{aligned}[t]
&(1+\alpha-\beta) + \frac{\alpha}{f^2 \cos^2 \Theta - \omega^2}( \grad \cdot \vec g_0 + \n_{\parallel} g_0 + g_0 \grad \cdot \vec e_{\parallel}) \\
&\quad+ \frac{\alpha}{f^2 \cos^2 \Theta - \omega^2} \frac{f^2}{\omega^2 -f^2} \lp - g_0 \cos \Theta \n_{\parallel}  \cos \Theta +  \cos \Theta \n_z g_0 - g_0 \cos^2 \Theta \n_{\parallel} g_0 \rp
\end{aligned}   \\
        &= 1+\alpha-\beta.
\end{align}

\section{The ray dynamics equations}
\label{sec:spherical}

In this section, the ray dynamics equations for axisymmetric rays are written using the spherical coordinates $[r,\theta]$
and the wave-vector components $[k_r, k_{\theta}]$ on the usual orthonormal basis $\vec{e}_r, \vec{e}_{\theta}$.
To derive them, it is convenient to start from the general Hamiltonian equations~\eq{Ham0} and \eq{Ham0bis} for spherical coordinates $[r,\theta]$. These equations are expressed in terms
of the covariant components of $\vec{k}$ on the natural basis $(k_r^{\rm nat}, k^{\rm nat}_{\theta})$, and a change of variables from $[r,\theta,k^{\rm nat}_r,k^{\rm nat}_{\theta}]$ to $[r,\theta,k_r,k_{\theta}]$
is necessary. From the relations $k_r = k_r^{\rm nat}$ and $k_{\theta} = k^{\rm nat}_{\theta}/r$,
Eqs.~\eq{Ham0} and \eq{Ham0bis} take the following form:
\beqa
\label{Ham1}
\frac{{\rm d} r}{{\rm d} t} & = & \frac{\partial H}{\partial k_r}, \\
\frac{{\rm d} \theta}{{\rm d} t} & = & \frac{1}{r} \frac{\partial H}{\partial k_{\theta}}, \label{Ham2}\\
\frac{{\rm d} k_r}{{\rm d} t} & = & - \frac{\partial H}{\partial r} + \frac{k_{\theta}}{r} \frac{\partial H}{\partial k_{\theta}},  \label{Ham3}\\
\frac{{\rm d} k_{\theta}}{{\rm d} t} & = & - \frac{1}{r} \frac{\partial H}{\partial \theta}  - \frac{k_{\theta}}{r} \frac{\partial H}{\partial k_r},    \label{Ham4}
\eeqa
where the Hamiltonian $H(r,\theta,k_r,k_{\theta})$ is given by the eikonal equation written in the form $\omega = H(r,\theta,k_r,k_{\theta})$.

To develop the previous equations, one possible way is to start from the original eikonal equation \eq{disp0} in the case $k_\phi=0$ written as
\beqa
\label{dispder}
(k^2+k_{\rm c}^2) \omega^2 = f^2 k_z^2 + N_0^2 k_{\perp}^2 + f^2\cos^2 \Theta k_{\rm c}^2,
\eeqa
and to calculate the partial derivatives of this expression with respect to the coordinates $[r,\theta,k_r,k_{\theta}]$.
To do so, one needs to express $k_z$ and $k_{\perp}$ as functions of $[r,\theta,k_r,k_{\theta}]$, namely
\beqa
\label{rel0}
k_z &= &\cos \theta \; k_r - \sin \theta \; k_{\theta}, \\
k_{\perp} &= &\sin(\theta - \Theta) k_r  + \cos(\theta - \Theta) k_{\theta},
\label{rel0bis}
\eeqa
where the last relation has been obtained from
\beqa
\label{rel1}
\vec{e}_{\parallel} &= &\cos \Theta \; \vec{e}_z + \sin \Theta \; \vec{e}_s, \\
\vec{e}_{\perp} &= & - \sin \Theta \; \vec{e}_z + \cos \Theta \; \vec{e}_s, \\
\vec{e}_{\perp} \cdot \vec{e}_r &=& \sin(\theta - \Theta), \\
\vec{e}_{\perp} \cdot \vec{e}_{\theta} &=& \cos(\theta - \Theta),
\eeqa
and $\vec e_s$ is the classical cylindrical unit vector pointing away from the $z$-axis.

The dynamical equations \eq{Ham1}--\eq{Ham4} in spherical coordinates are
 \begin{align}
\frac{{\rm d} r}{{\rm d} t}   &=  \frac{1}{\omega(k^2+k_{\rm c}^2)} \left\lbrace k_r \left[{N_0}^2 \sin^2(\theta-\Theta)+f^2\cos^2\theta-\omega^2\right]+k_\theta\left[{N_0}^2 \sin(\theta-\Theta) \cos(\theta-\Theta)-f^2 \sin\theta \cos\theta\right] \right\rbrace,    \\
\frac{{\rm d} \theta}{{\rm d} t}    &=  \frac{1}{\omega r (k^2+k_{\rm c}^2)} \left\lbrace k_r \left[{N_0}^2 \sin(\theta-\Theta) \cos(\theta-\Theta)-f^2 \sin\theta \cos\theta\right]+k_\theta \left[{N_0}^2 \cos^2(\theta-\Theta)+f^2\sin^2\theta-\omega^2\right] \right\rbrace,    \\
\frac{{\rm d} k_r}{{\rm d} t}    &=
\begin{aligned}[t]
  &\frac{1}{\omega(k^2+k_{\rm c}^2)} \bigg \lbrace
                k_r\frac{k_\theta}{r} \left[{N_0}^2 \sin(\theta-\Theta) \cos(\theta-\Theta)-f^2 \sin\theta \cos\theta\right]+\frac{{k_\theta}^2}{r} \left[{N_0}^2 \cos^2(\theta-\Theta)+f^2\sin^2\theta-\omega^2\right] \\
                & \quad +   \frac{\partial_r({k_{\rm c}}^2)}{2} (\omega^2-f^2 \cos^2\Theta)-\frac{\partial_r({N_0}^2)}{2} \left[k_r \sin(\theta-\Theta)+k_\theta \cos(\theta-\Theta)\right]^2\\
                & \quad +    \partial_r\Theta\left({N_0}^2 \left\lbrace\sin(\theta-\Theta) \cos(\theta-\Theta) ({k_r}^2-{k_\theta}^2)+k_r k_\theta \left[2 \cos^2(\theta-\Theta)-1\right]\right\rbrace+f^2 \sin\Theta \cos\Theta {k_{\rm c}}^2\right) \bigg \rbrace,
\end{aligned} \\
        \frac{{\rm d} k_{\theta}}{{\rm d} t}    &=
\begin{aligned}[t]
  &\frac{1}{r\omega(k^2+k_{\rm c}^2)} \bigg \lbrace
                    f^2 \left[\sin\theta \cos\theta ({k_r}^2-{k_\theta}^2)+k_r k_\theta (2 \cos^2\theta-1)\right]+\frac{\partial_\theta({k_{\rm c}}^2)}{2} (\omega^2-f^2 \cos^2\Theta) \\
                & \quad -   \frac{\partial_\theta({N_0}^2)}{2} \left[(k_r \sin(\theta-\Theta)+k_\theta \cos(\theta-\Theta)\right]^2+f^2 \sin\Theta \cos\Theta {k_{\rm c}}^2 \partial_\theta\Theta\\
                & \quad +    (\partial_\theta\Theta-1) \left\lbrace\sin(\theta-\Theta) \cos(\theta-\Theta) ({k_r}^2-{k_\theta}^2)+k_r k_\theta \left[2 \cos^2(\theta-\Theta)-1\right]\right\rbrace {N_0}^2  \\
                & \quad - k_\theta k_r  \left[{N_0}^2 \sin^2(\theta-\Theta)+f^2\cos^2\theta-\omega^2\right] - k_\theta^2 \left[{N_0}^2 \sin(\theta-\Theta) \cos(\theta-\Theta)-f^2 \sin\theta \cos\theta\right]  \bigg \rbrace,
\end{aligned}
\end{align}
These equations have been used for the numerical computation of the gravito-inertial dynamics in barotropic models of uniformly rotating stars.

\section{Derivation of relations \eq{Perp}, \eq{Retour}, and \eq{Retourbis}}
\label{sec:surface}

It is also useful to derive other relations to understand ray dynamics. Below, the relation \eq{Perp} between the group velocity and the phase velocity is derived.
Using the first two equations of the Hamiltonian system \eq{Ham1} and \eq{Ham2}, we have
 \beqa
\frac{{\rm d} \vec{x}}{{\rm d}t} \cdot \vec{k} = k_r \frac{\partial H}{\partial k_r}  + k_{\theta} \frac{\partial H}{\partial k_{\theta}}.
\eeqa
Then, taking the derivatives of the eikonal equation \eq{dispder} with respect to $k_r$ and $k_\theta$, we obtain
\beqa
\label{partial0}
(k^2+k_{\rm c}^2)\omega \frac{\partial H}{\partial k_r} + \omega^2 k_r = f^2 k_z  \frac{\partial k_z}{\partial k_r}  + N_0^2 k_{\perp} \frac{\partial k_{\perp}}{\partial k_r}, \\
(k^2+k_{\rm c}^2)\omega \frac{\partial H}{\partial k_\theta} + \omega^2 k_\theta = f^2 k_z  \frac{\partial k_z}{\partial k_\theta}  + N_0^2 k_{\perp} \frac{\partial k_{\perp}}{\partial k_\theta}.\label{partial0_end}
\eeqa
It follows that
 \beqa
\omega (k^2+k_{\rm c}^2) \frac{{\rm d} \vec{x}}{{\rm d}t} \cdot \vec{k} = -\omega^2 k^2 + f^2 k_z\left(k_r \frac{\partial k_z}{\partial k_r} + k_\theta \frac{\partial k_z}{\partial k_\theta}\right) + N_0^2 k_\perp
\left(k_r \frac{\partial k_{\perp}}{\partial k_r} + k_\theta \frac{\partial k_{\perp}}{\partial k_\theta}\right).
\eeqa
Then,
from Eqs.~\eq{rel0} and \eq{rel0bis}, we have
\beqa
\label{partial1}
\frac{\partial k_z}{\partial k_r} &=& \cos \theta, \\
\frac{\partial k_z}{\partial k_\theta} &=& -\sin \theta, \\
\frac{\partial k_{\perp}}{\partial k_r} &=& \sin(\theta-\Theta),                \\
\frac{\partial k_{\perp}}{\partial k_\theta} &=& \cos(\theta-\Theta),\label{partial1_end}
\eeqa
and from there we deduce that
 \beqa
\omega (k^2+k_{\rm c}^2) \frac{{\rm d} \vec{x}}{{\rm d}t} \cdot \vec{k} & =& -\omega^2 k^2 + f^2 k^2_z + N_0^2 k^2_\perp \\
&= & (\omega^2 - f^2 \cos^2 \Theta) k^2_{\rm c},
\label{Perp_an}
\eeqa
which corresponds to relation \eq{Perp}.

Similar calculations enable us to derive the expression \eq{Retour}. Indeed, using the first two equations of the Hamiltonian system \eq{Ham1} and \eq{Ham2} and the expression
of $\vec{e}_{\parallel}$ in Eq.~\eq{rel1}, we have
\beqa
\frac{{\rm d} \vec{x}}{{\rm d}t} \cdot \vec{e}_{\parallel}  = \cos(\theta-\Theta) \frac{\partial H}{\partial k_r} - \sin(\theta-\Theta) \frac{\partial H}{\partial k_\theta}.
\eeqa
Then, from Eqs.~\eq{partial0}, \eq{partial0_end}, and \eq{partial1}--\eq{partial1_end}, we find
\beqa
\label{boubou}
\omega (k^2+k_{\rm c}^2) \frac{{\rm d} \vec{x}}{{\rm d}t} \cdot \vec{e}_{\parallel}  = f^2 k_z \cos \Theta - \omega^2 k_{\parallel}.
\eeqa
Solving the eikonal equation \eq{disp1} as a quadratic equation for the variable $k_{\parallel}$, we find
\beqa
k_{\parallel}^{\pm} = \frac{f^2 \cos \Theta \sin \Theta k_\perp \pm \sqrt{\delta}}{f^2 \cos^2 \Theta - \omega^2},
\eeqa
where $\delta$ has been already defined by Eq.~\eq{del}.
Then,  the relation \eq{Retour} is obtained replacing $k_z$ by $k_z = k_{\parallel} \cos \Theta - k_{\perp} \sin \Theta$ in Eq.~\eq{boubou}.
From the difference of Eq.~\eq{Perp} and Eq.~\eq{Retour}, an expression for ${\rm d} \vec{x}/{\rm d} t \cdot \vec e_{\perp} $
is obtained. Using the above expression of $k_{\parallel}^{\pm}$ and Eq.~\eq{del}, we derive Eq.~\eq{Retourbis}. 

\section{Test of the numerical method on a regular cusp point}
\label{sec:test}

We considered the simple case of gravity rays in a plane-parallel atmosphere characterised by an exponential variation of the Brunt-V\"{a}is\"{a}l\"{a} frequency in the $z$ direction.
The ray dynamics equations derived from the dispersion relation $\omega=N_0 k_x/\sqrt{k_x^2+k_z^2}$ are
\beqa
\label{dxdt}
\frac{{\rm d}x}{{\rm d}t}=\frac{\omega}{k_x}\frac{{k_z}^2}{{k_x}^2+{k_z}^2}, \\
\label{dzdt}
\frac{{\rm d}z}{{\rm d}t}=-\omega\frac{k_z}{{k_x}^2+{k_z}^2},
\eeqa
where $k_x$ is constant, $k_z=k_{z0}-\gamma\omega t$, and $\gamma={\rm d}\ln N_0/{\rm d}z$.
An analytical solution of these equations going through a regular cusp point of coordinates $(x_{\rm r}, z_{\rm r})$ is
\beqa
\left\lbrace
\begin{array}{ll}       
x(u)    &=      x_{\rm r}-\frac{1}{\gamma}\left(u-\arctan u\right),   \\
z(u)    &=      z_{\rm r}+\frac{1}{2\gamma}\ln(1+u^2),\\
\end{array}
\right.
\eeqa
where the time variable $t$ has been replaced by $u = k_z/k_x$.
The cusp point is reached at $u=0$.
The code used in the present paper was able to reproduce this analytical solution with a controlled level of
accuracy.

\section{Unstable fixed point of the non-rotating radial ray dynamics}
\label{sec:hyperbolic}

In non-rotating stars, the ray dynamics equations can be solved separately in the radial and latitudinal coordinates.
The radial dynamics is governed by Eq.~(\ref{rot0}). For some values of the frequency, the phase portrait of the radial dynamics
is typical of double-well-potential systems with two elliptic regions around the two fixed points at the minima of the potential separated
from high-energy motions by a separatrix that goes through the unstable hyperbolic fixed point at the maximum of the potential. This is illustrated
by panel (a) of Fig.~\ref{SdPsep}. Equation~(\ref{rot0}) can be seen as a classical Hamiltonian $H_r = 1/2 k_r^2 + V_r$ with $H_r=0$ and
$V_r$ a potential
that depends on two parameters, $\omega $ and $L$. In our case, the separatrix separates low-$L$ from high-$L$ motions.
When such a system is moved away
from integrability, chaos appears first near the unstable hyperbolic point.
Here, we show that using an envelope model we can analytically determine 
the frequency range where the non-rotating radial dynamics shows a double-well-potential behaviour
as well as the separatrix and the position of the unstable point.

In envelope models of stars, the radial dependence of the mass inside a given radius is neglected.
Within this approximation,
the hydrostatic equation can be integrated to provide simple expressions for the profile
of the thermodynamic quantities \citep{C68}. In particular, the Brunt-V\"{a}is\"{a}l\"{a} frequency can be written as
\begin{equation}
N_0^2 =  \frac{\alpha(\mu+1)}{\Gamma_1} \frac{G M}{R^3} \frac{R^4}{r^3(R-r)},
\label{No}
\end{equation}
and can be shown to possess a local minimum $N_{0,{\rm min}}$ at $r/R= 0.75$. This local minimum is
expected to be a generic feature of radiative stellar envelopes.
It is at the origin of the existence of an unstable fixed point of the ray dynamics.

Using Eq.~\eq{No}, Eq.~(\ref{rot0}) governing the radial dynamics becomes
\begin{equation}
R^2 k_r^2 = \left( \frac{a}{\tilde{\omega}^2} \frac{1}{x^3(1-x)} - 1 \right) \frac{L^2}{x^2} - b \frac{1}{x^2(1-x)^2},
\end{equation}
where we have introduced the dimensionless quantities $\tilde{\omega} = \omega/\sqrt{G M/R^3}$, $x=r/R$ and the notations $a = \alpha(\mu+1)/\Gamma_1$ and $b=(\mu+1)(\mu-1)/4$.
In our one-dimensional system, $k_r$ and the radial derivative of $V_r$ must vanish at fixed points. As $H_r =0$, the first condition
implies that $V_r$ also vanishes at the fixed point. The equilibrium will be unstable if $V_r$ is maximum there.
Asking that both $V_r$ and its radial derivative vanish is equivalent to solve the following system
of equations for the variable $x$ excluding the $x=0$ case:
\beqa
\left\lbrace
\begin{array}{ll}
\dfrac{a}{\tilde{\omega}^2} (1-x) - \dfrac{b}{L^2} x^3 -x^3 (1-x)^2 & = 0,\\
\dfrac{a}{\tilde{\omega}^2} (1-x) (4x-3) - 2 \dfrac{b}{L^2} x^4 & = 0.
\end{array}
\right.
\eeqa

Using two new variables $u=1-x$ and $v=x^3$, the system above is equivalent to solving a cubic equation for $u$, a second equation giving the relation between
$\tilde{\omega}$ and $L$ as follows:
\beqa
\left\lbrace
\begin{array}{l}
u^3 -\dfrac{u^2}{4} + \dfrac{b}{2 L^2} u + \dfrac{b}{4 L^2} =0, \\
\tilde{\omega}^2 = a \dfrac{L^2}{2b} \dfrac{u(1-4u)}{(1-u)^4}.
\end{array}
\right.
\eeqa

The cubic equation has a unique real solution if $L$ is smaller than a minimum value $L_{\rm min}$, and this solution is not physical as it gives a point outside the star.
Above $L_{\rm min}$, there are in addition two real solutions among
which
the smaller solution
corresponds to the unstable point and the larger solution to the outer stable point.
Simple analytical solutions are obtained for the two extreme cases $L=L_{\rm min}$ and $L = +\infty$.

In conclusion,
an unstable hyperbolic point is present in the frequency range $0.9348 N_{0,{\rm min}} \leq \omega \leq N_{0,{\rm min}}$. As $\omega$ increases in this interval,
the position of the unstable
point decreases from $r = 0.8406 R$ to $r= 0.75 R$, while the $L$ value of the separatrix increases from
$L_{\rm min} =\sqrt{\frac{16(\mu+1)(\mu-1)}{19\sqrt{57}-143}}$ to $L = +\infty$.
It is interesting that some properties of the unstable hyperbolic point, namely the radius and the frequency interval normalised
by $N_{0,{\rm min}}$, do not depend on the polytropic index. For a $\mu=3$ polytropic model, $L_{\rm min} = 16.92$.

We compared these analytical solutions to numerical determinations of the hyperbolic point obtained for the full $\mu=3$ polytropic model,
that is without the envelope approximation. The agreement is fairly good since the numerical solution gives $c_2 N_{0,{\rm min}} \leq \omega \leq N_{0,{\rm min}}$
with $0.9371 <c_2< 0.9372$ and $16.80 < L_{\rm min}< 16.81$.

\end{document}